\documentclass[prl,aps,floats,superscriptaddress,twocolumn]{revtex4-1}

\usepackage[utf8]{inputenc}
\usepackage{amsmath,amssymb,amsbsy,bm,ulem}
\usepackage{dsfont,verbatim,varwidth,dcolumn}
\usepackage{graphicx,epsfig,epstopdf,subfigure,float}
\usepackage[usenames,dvipsnames]{xcolor}
\usepackage{subfiles}

\def\ua{{\uparrow}}
\def\da{{\downarrow}}
\newcommand{\vk}{\vec{k}}
\newcommand{\vK}{\vec{K}}
\newcommand{\STO}{SrTiO$_3$}
\newcommand{\LAO}{LaAlO$_3$}

\makeatother

\begin{document}

\title{Symmetry and correlation effects on band structure explain the anomalous transport properties of $(111)$ LaAlO$_3$/SrTiO$_3$}
\author{Udit Khanna}
\affiliation{Raymond and Beverly Sackler School of Physics and Astronomy, Tel-Aviv University, Tel Aviv, 6997801, Israel}
\author{P. K. Rout}
\affiliation{Raymond and Beverly Sackler School of Physics and Astronomy, Tel-Aviv University, Tel Aviv, 6997801, Israel}
\author{Michael Mograbi}
\affiliation{Raymond and Beverly Sackler School of Physics and Astronomy, Tel-Aviv University, Tel Aviv, 6997801, Israel}
\author{Gal Tuvia}
\affiliation{Raymond and Beverly Sackler School of Physics and Astronomy, Tel-Aviv University, Tel Aviv, 6997801, Israel}
\author{Inge Leermakers}
\affiliation{High Field Magnet Laboratory (HFML-EFML), Radboud University, 6525 ED Nijmegen, The Netherlands}
\author{Uli Zeitler}
\affiliation{High Field Magnet Laboratory (HFML-EFML), Radboud University, 6525 ED Nijmegen, The Netherlands}
\author{Yoram Dagan}
\affiliation{Raymond and Beverly Sackler School of Physics and Astronomy, Tel-Aviv University, Tel Aviv, 6997801, Israel}
\author{Moshe Goldstein}
\affiliation{Raymond and Beverly Sackler School of Physics and Astronomy, Tel-Aviv University, Tel Aviv, 6997801, Israel}

\date{\today}

\begin{abstract}
  The interface between the two insulating oxides SrTiO$_3$ and LaAlO$_3$ gives rise to a
  two-dimensional electron system with intriguing transport phenomena, including superconductivity,
  which are controllable by a gate. Previous measurements on the $(001)$ interface have shown that the superconducting
  critical temperature, the Hall density, and the frequency of quantum oscillations, vary
  nonmonotonically and in a correlated fashion with the gate voltage. In this paper we experimentally 
  demonstrate that the $(111)$ interface features a qualitatively distinct behavior, in which the 
  frequency of Shubnikov-de Haas oscillations changes
  monotonically, while the variation of other properties is nonmonotonic albeit uncorrelated.
  We develop a theoretical model, incorporating the different symmetries of these interfaces as well as 
  electronic-correlation-induced band competition. We show that the latter dominates at $(001)$, leading to similar 
  nonmonotonicity in all observables, while the former is more important at $(111)$, giving rise to highly 
  curved Fermi contours, and accounting for all its anomalous transport measurements.
\end{abstract}

\maketitle

\textit{Introduction.---}
The high-mobility two-dimensional electron system (2DES) at the interface of SrTiO$_3$ and
LaAlO$_3$ \cite{hwang2004} shows a variety of quantum transport phenomena 
\cite{dagan2010,CavigliaSdH,mbs2010,maniv100,pryds}, in addition to a rich phase
diagram including magnetism \cite{brinkman,beena2011,Yoram2014} and superconductivity
\cite{Reyren1196,caviglia2008,bell2009} at low temperatures. The multi-orbital band structure of
the system, which gives rise to this physics, has been the subject of many studies. The electronic
structure of the interface has been probed via optical methods such as X-ray absorption spectroscopy
\cite{XLD100,Herranz2014} and angle resolved photo-emission spectroscopy \cite{ARPES1,ARPES2} as
well as through magnetotransport \cite{joshua2012,ruhman2014,MRS100,maniv100,smink100}, which were supplemented 
by density functional theory based \textit{ab-initio} calculations \cite{santander2011,DFT0,DFT1,DFT2,DFT3} and 
analytical studies~\cite{michaeli,randeria}. Most studies concentrated on the $(001)$ interface, although a 
conducting 2DES can arise in other interfaces~\cite{herranz2012}. This has changed recently with several works
\cite{mckeown,santander2014,cooper17,YoramAMR111,yoramSOSC,yoram1805,venkatesan,caviglia111,Nematic111,
venkatesan18,Satoshi11,Satoshi13,Satoshi18,111DFT1,111DFT2,xld111} indicating that the $(111)$ interface has a
distinct electronic structure with novel properties.

To elucidate the electronic properties of $(111)$ \LAO/\STO, we embarked on a combined experimental and theoretical study.
Experimentally we focus on magnetotransport at the $(111)$ interface (Hall effect, quantum oscillations, and 
superconductivity), which shows surprising differences from the $(001)$ interface \cite{maniv100}: In $(001)$ all 
these quantities are nonmonotonic and reach their maximum at roughly the same gate voltage, whereas at $(111)$ the 
quantum oscillations frequency is monotonic, and the peaks in the Hall density and superconducting transition temperature are well-separated.
To understand these results, we  calculate the correlation-induced band structure of the 2DES, taking into account the crystal
structure and the change in symmetry from the bulk (octahedral) to the interface [triangular in $(111)$, square in $(001)$]
\cite{111DFT1,111DFT2,xld111}. We elucidate the different behavior of the $(111)$ as compared to the $(001)$ 
interface: While the latter is dominated by interaction-induced population transfer, the former is governed by 
symmetry-induced Fermi contour shape.
The resulting transport coefficients nicely follow the experimental data. 

\begin{figure}
\includegraphics[width=0.9\columnwidth]{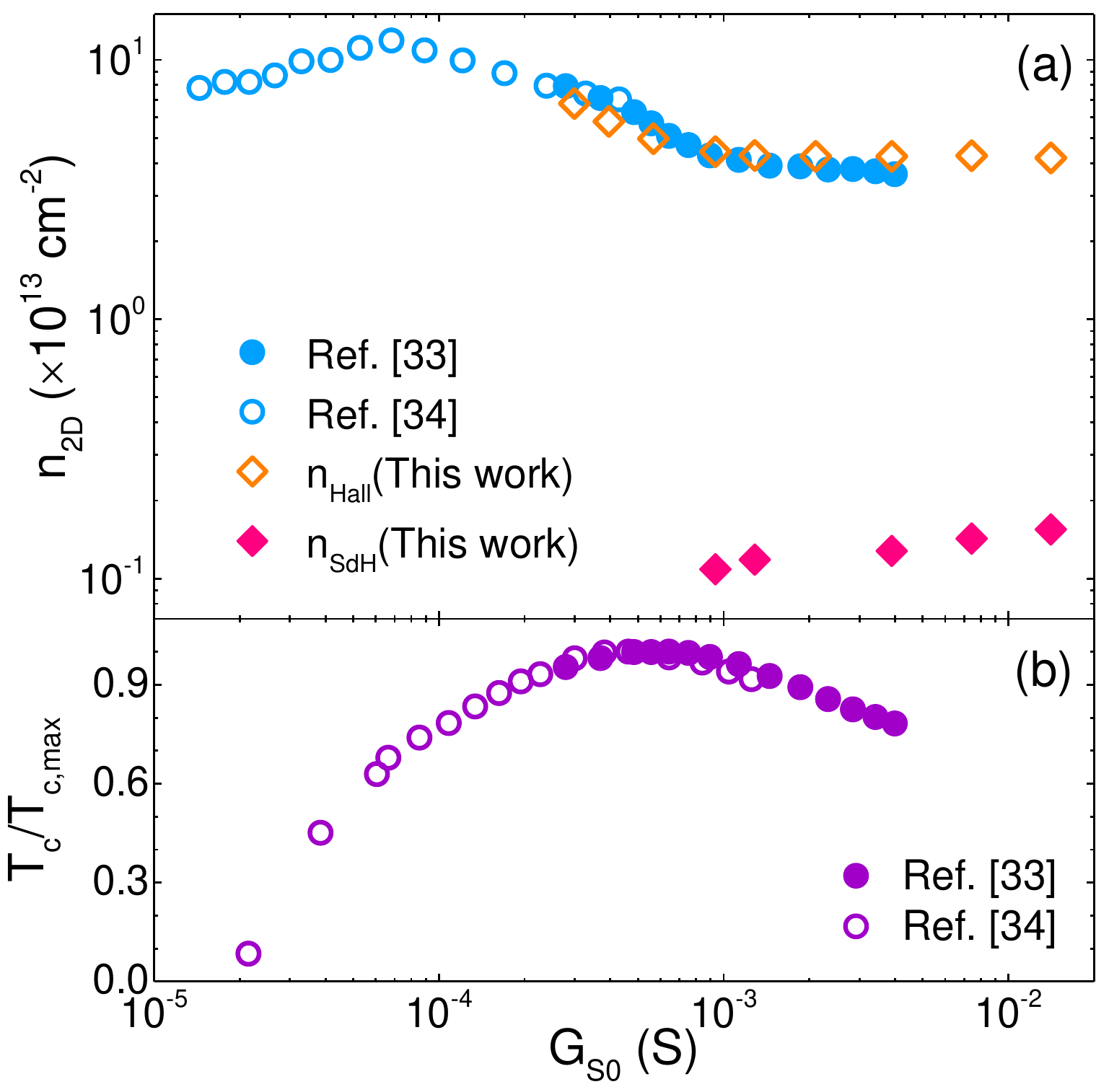}
  \caption{Gate dependence of transport parameters. (a) The sheet carrier density ($n_{\text{2D}}$)
  determined from low-field Hall measurements and quantum oscillations as a function of the zero
  field sheet conductance ($G_{\text{S0}}$). (b) Normalized superconducting
  critical temperature $T_{\text{c}}/T_{\text{c,max}}$ as a function of $G_{\text{S0}}$.}
\label{fig-expt}
\end{figure}

{\it Transport measurements.---}
$14$ monolayer thick epitaxial thin film of \LAO~were grown on atomically flat Ti-terminated
\STO~$(111)$ substrate using the pulsed laser deposition technique in combination with
reflection high energy electron diffraction. Details of the deposition procedure and
substrate treatment are described in Ref.~\cite{YoramAMR111}. Electrical transport measurements
of the $80$~$\mu\text{m} \times 260$~$\mu\text{m}$ Hall bar, patterned along the {[}1$\bar{2}$1{]}
direction using optical lithography \cite{YoramAMR111}, were performed by a four probe ac technique with a
current of $50$ nA in a custom made ${}^{3}$He cryostat equipped with a $35$ T magnet.

We investigated magnetotransport at the $(111)$ interface under a perpendicular field to understand the
behavior of the carrier density ($n_{\text{2D}}$) as a function of temperature ($T$) and gate
voltage ($V_{\text{g}}$) in a back-gated device. $n_{\text{2D}}$ was extracted using both the
Hall density [$n_{\text{Hall}} = (e {R_\text{H}})^{-1}$, where $R_{\text{H}}$ is the slope of the
low-field Hall resistivity] and the Shubnikov-de Haas (SdH) oscillations (through the Onsager relation
\cite{supp}) observed at higher magnetic fields. We also studied corresponding variation of the superconducting 
transition temperature. The back-gate was employed to control the carrier density
and vary the sheet conductance ($G_{\text{S}}$). Since the gate response changes between different
sample cool-downs and $V_{\text{g}}$ sweeps, we present the results in Fig.~\ref{fig-expt} as a function
of the zero field conductance $G_{\text{S0}}$ \cite{yoram1805}. The $V_{\text{g}}$ dependence
can be found in~\cite{supp}.

\begin{figure}[t]
\includegraphics[width=\columnwidth]{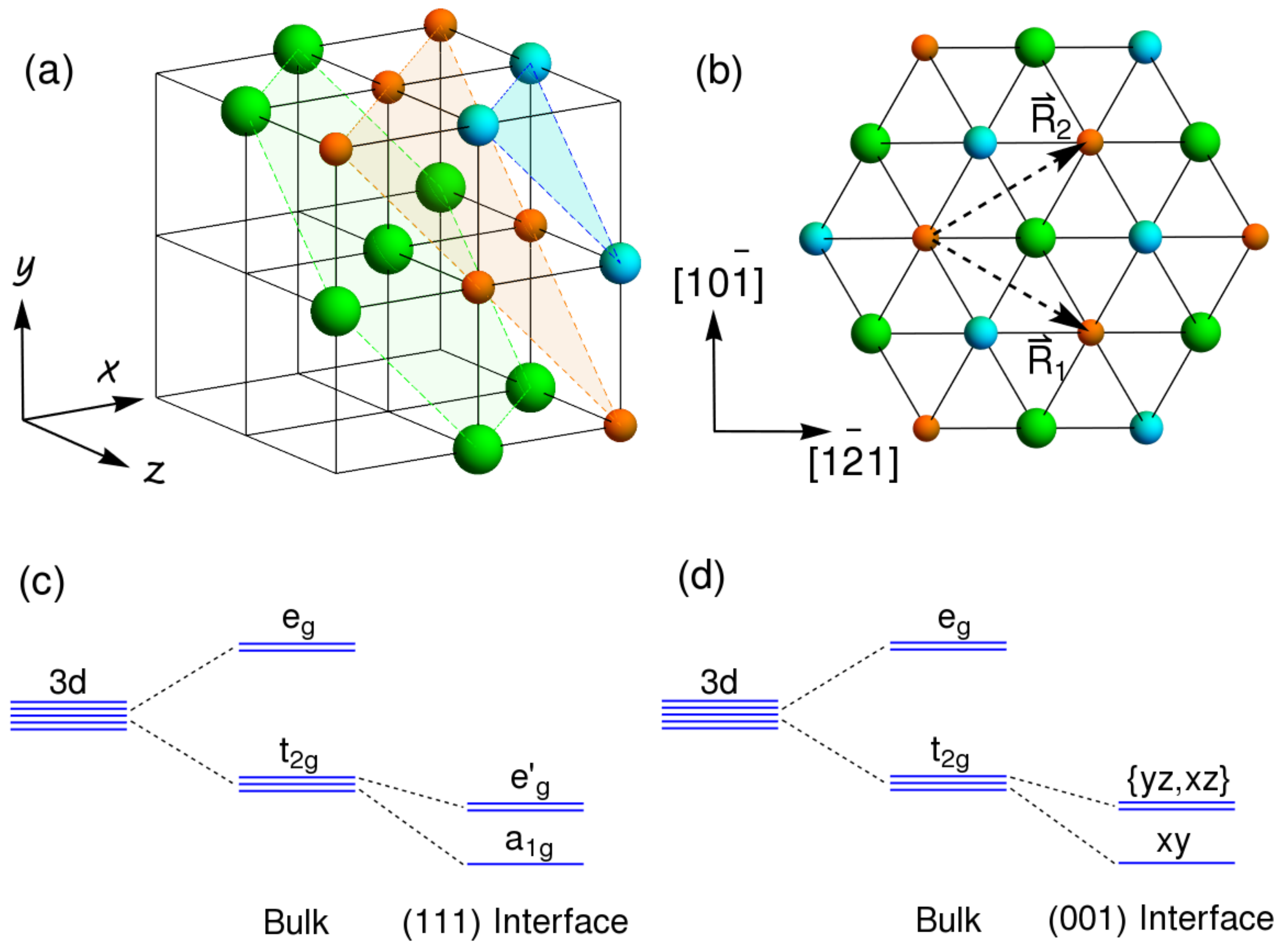}
\caption{Crystal structure of the $(111)$ \LAO/\STO~interface. (a) Three inequivalent layers of Ti atoms
  ($\text{green } \equiv \text{Ti}_{1} \text{, orange} \equiv \text{Ti}_{2} \text{, blue} \equiv \text{Ti}_{3}$)
  forming one unit cell at the $(111)$ interface of a cubic lattice.
  (b) Top view of the $(111)$ trilayer. $\vec{R}_{1,2}$ are the lattice vectors of the triangular
  lattice at the interface. (c),(d) Level structure of the Ti 3d-orbitals at the $(111)$ and $(001)$
  interfaces, respectively. The bulk cubic structure leads to the splitting into t$_{\text{2g}}$
  and e$_{\text{g}}$ orbitals. (c) The trigonal crystal symmetry at the $(111)$ interface
  leads to further splitting of t$_{\text{2g}}$ into a$_{\text{1g}}$ and e$_\text{g}'$ orbitals.
  (d) The in-plane crystal symmetry does not change at the $(001)$ interface but the surface
  confinement lifts the degeneracy.}
\label{fig-a1g}
\end{figure}

Fig.~\ref{fig-expt}(a) compares the variation of carrier density from the SdH analysis
($n_{\text{SdH}}$) and the Hall measurement ($n_{\text{Hall}}$) with the gate voltage
($V_{\text{g}}$), while Fig.~\ref{fig-expt}(b) presents the corresponding dependence of the superconducting
critical temperature ($T_\text{c}$). 
The observed variation and values of $n_{\text{Hall}}$ are
consistent with our previous results on the $(111)$ interface \cite{yoram1805,yoramSOSC}
[also shown in Fig.~\ref{fig-expt}(a)] and with other recent results \cite{venkatesan,caviglia111}.

Curiously, we find that while $n_{\text{Hall}}$ and $T_{\text{c}}$ are non-monotonic functions
of $V_\text{g}$, $n_{\text{SdH}}$ changes monotonically. Moreover, the peak in $n_{\text{Hall}}$ appears
when quantum oscillations are not observable. These features are strikingly different from our previous
measurements on the $(001)$ interface \cite{maniv100}. In the latter case, $n_{\text{SdH}}$
also changes non-monotonically with $V_\text{g}$ and the maximal $n_{\text{SdH}}$, $n_{\text{Hall}}$, and 
$T_{\text{c}}$ appear at roughly the same gate voltage.

At both interfaces $n_{\text{SdH}}$ is much smaller than $n_{\text{Hall}}$. Since the
SdH signal decays exponentially with inverse scattering time, this indicates the presence of
two low-energy bands in the electronic structure with different mobilities. Therefore, both bands
would contribute to $n_{\text{Hall}}$ but only the mobile one would be observable through
the quantum oscillation measurements.

We note that the band structure of $(111)$ \LAO/\STO~has recently been probed using Hall measurements~\cite{caviglia111}. 
However, the Hall coefficient receives contributions from all the bands and also depends on the corresponding 
scattering times, making it hard to decipher the band structure.
The crucial new ingredient here is the quantum oscillations, which directly probe the population of the more 
mobile band, and demonstrate the qualitative difference between the $(111)$ and $(001)$ interfaces. 
These allow us to develop a complete and consistent theoretical picture for both interfaces, as we now turn to describe.

{\it Theoretical Model.---}
We first consider the orbital character of the relevant levels at the two interfaces.
\textit{Ab-initio} studies \cite{santander2011,DFT0,DFT1} show that the low energy conduction bands
in bulk \STO\ are composed of the t$_{\text{2g}}$ orbitals of the Ti atoms. These are degenerate in the bulk due to their cubic
arrangement (the low temperature structural distortions are negligible for our purposes), which imparts octahedral
symmetry to the band structure. However, the reduced symmetry at the interfaces can lift the
degeneracy and modify the orbital character.

At the $(001)$ interface, Ti atoms form a square lattice, which does not modify the in-plane
crystal-field. In combination with the confining potential, the degeneracy of the t$_{\text{2g}}$
orbitals is lifted but the orbital character is not modified. Specifically, if the confinement is along the
$z$ direction, then the $xy$ orbital is lowered in energy due to its higher effective mass in the confinement direction \cite{santander2011}
[Fig.~\ref{fig-a1g}(d)].
On the other hand, at the $(111)$ interface, Ti atoms form a stacked triangular lattice with
three interlaced layers [Fig.~\ref{fig-a1g}(a),(b)]. This changes the bulk octahedral symmetry
to triangular at the interface and introduces a new {\it in-plane} crystal field \cite{khomskii},
which hybridizes the t$_{\text{2g}}$ orbitals, forming $|a_{\text{1g}}\rangle = (|xy\rangle +
|yz\rangle + |xz\rangle)/\sqrt{3}$ and $|e_{\text{g}\pm}'\rangle = (|xy\rangle +
\omega^{\pm 1}|yz\rangle + \omega^{\pm 2} |xz\rangle)/\sqrt{3}$ where $\omega = e^{2\pi i /3}$.
Their splitting is sensitive to details of the interface. Here, we choose
parameters such that a$_{\text{1g}}$ is lower in energy [Fig.~\ref{fig-a1g}(c)], 
in accordance with recent XLD experiments \cite{xld111} and DFT calculations \cite{111DFT1,111DFT2}. 

Next, we employ a tight-binding model with these orbitals on the first three inequivalent layers
[Fig.~\ref{fig-a1g}(a)], keeping track of the separation and connectivity of sites
on the different layers \cite{fnLayers}. In the basis, $\{ |a_{\text{1g}}\rangle, |e^\prime_{\text{g}+}\rangle,
|e^\prime_{\text{g}-}\rangle \} \otimes \{ |\text{Ti}_{1}\rangle, |\text{Ti}_{2}\rangle, |\text{Ti}_{3}\rangle \}
\otimes \{|\ua\rangle, |\da\rangle \}$, the hopping terms can be written as $18 \times 18$
matrices given by
\begin{align}
  &H^{(111)}_{\text{0}}(\vK) =  \left( \begin{array}{ccc}
    A(\vK) & B(\vK) & B^{\dagger}(\vK) \\
    B^{\dagger}(\vK) & A(\vK) & B(\vK) \\
    B(\vK) & B^{\dagger}(\vK) & A(\vK)
  \end{array} \right) \otimes \mathcal{I}_2 ,
\end{align}
where, the block matrices $A(\vK)$ and $B(\vK)$ are,
\begin{widetext}
\begin{align}
  A(\vK) &= -\frac{(2t+t^\prime)}{3} \left( \begin{array}{ccc}
      0 & e^{-i K_2} f_{0}(\vK) & 0 \\
      e^{i K_2} f_{0}(-\vK) & 0 & e^{-i K_1} f_{0}(\vK) \\
      0 & e^{i K_1} f_{0}(-\vK) & 0 \end{array} \right) -
      \frac{t^{\prime\prime}}{3} \left( \begin{array}{ccc}
      2 \epsilon_{0}(\vK) & 0 & f_{0}(-\vK) \\
      0 & 2 \epsilon_{0}(\vK) & 0 \\
      f_{0}(\vK) & 0 & 2 \epsilon_{0}(\vK) \end{array} \right), \\
  B(\vK) &= \omega^2 \frac{(t-t^\prime)}{3} \left( \begin{array}{ccc}
      0 & e^{-i K_2} f_{\omega}(\vK) & 0 \\
      e^{i K_2} f_{\omega}(-\vK) & 0 & e^{-i K_1} f_{\omega}(\vK) \\
      0 & e^{i K_1} f_{\omega}(-\vK) & 0 \end{array} \right) -
      \omega^2 \frac{t^{\prime\prime}}{3} \left( \begin{array}{ccc}
      2 \epsilon_{\omega}(\vK) & 0 & f_{\omega}(-\vK) \\
      0 & 2 \epsilon_{\omega}(\vK) & 0 \\
      f_{\omega}(\vK) & 0 & 2 \epsilon_{\omega}(\vK) \end{array} \right),
\end{align}
\end{widetext}
where, $t$ and $t^\prime$ are the light and heavy nearest neighbor hopping
amplitudes while $t^{\prime\prime}$ is the next-nearest neighbor hopping.
$f_{0}(\vK) = 1 + e^{i K_1} + e^{i K_2}$, $f_{\omega}(\vK) = 1 + \omega e^{i K_1} + \omega^2 e^{i K_2}$,
$\epsilon_{0} (\vK) = \cos(K_1) + \cos(K_2) + \cos(K_1 - K_2)$ and
$\epsilon_{\omega} (\vK) =  \cos(K_1 - K_2) + \omega \cos(K_2) + \omega^2 \cos(K_1) $, where
$K_{1,2} \in [-\pi,\pi]$,
the two-dimensional Brillouin zone.
The atomic spin-orbit coupling is an on-site term mixing the orbitals
and spin states. Taking the spin quantization axis along
the $(111)$ direction, the spin-orbit coupling is,
\begin{align}
  &H^{(111)}_{\text{SO}} = \frac{\Delta_{\text{so}}}{2} \left( \begin{array}{ccc}
    0 & -\sqrt{2}\sigma^{+} & \sqrt{2}\sigma^{-} \\
     -\sqrt{2}\sigma^{-} & - \sigma_{z} & 0 \\
     \sqrt{2}\sigma^{+} & 0 & \sigma_{z}
  \end{array} \right),
\end{align}
where $\sigma^\pm = (\sigma_x \pm i \sigma_y)/2$, with $\sigma_{x,y,z}$ being the Pauli matrices.
Additionally, the single-particle Hamiltonian includes the trigonal crystal-field $\Delta_{\text{cf}}$
(which lifts the degeneracy between the orbitals) and a linear confining potential $V_{\text{c}}$
(which lifts the layer degeneracy)~\cite{supp}.

Finally, correlation effects are incorporated through an on-site Hubbard term
$\sum_{\text{r}} \sum_{\text{I} \neq \text{J}} U n_{\text{rI}} n_{\text{rJ}}$, which includes both inter-orbital
and intra-orbital repulsion (assumed to be of equal strength in order to reduce the number of free parameters).
The two-body term is then treated in the Hartree-Fock approximation. The mean-field ansatz is that the ground
state is invariant under time-reversal and has the full symmetry of the (interface) crystal structure, \textit{i.e.}
the C$_{3v}$ group at the $(111)$ interface (we have verified that tetragonal distortions etc. have a small effect on our results).
Under this assumption, the Hubbard term reduces to a one-body term with four independent real parameters (per layer) ---
the occupancy of the three orbitals (which appear in the Hartree terms) and one spin-mixing average (Fock term)
which renormalizes the spin-orbit interaction.
We note that a state with the full crystal symmetry must have equal occupancy of the $xy$, $yz$ and $xz$ orbitals.
Therefore, in terms of the original t$_{\text{2g}}$ orbitals, there is only one independent Hartree term and three Fock terms.

The mean-field Hamiltonian is solved self-consistently, using $t = V_{\text{c}} = 437.5$ meV, $t' = t'' = 20$ meV,
$\Delta_{\text{so}} = 3$ meV, $\Delta_{\text{cf}} = 2$ meV and $U = 2$ eV. These parameter values are close to those
employed previously for the $(001)$ interface \cite{joshua2012,ruhman2014,maniv100,MRS100}. Although surface
reconstruction can lead to different parameters at the two interfaces, the qualitative behavior is not expected
to change.

{\it Theoretical Results.---}
Fig.~\ref{Fig-theo} shows the results of the self-consistent calculation for the $(111)$ interface.
Figs.~\ref{Fig-theo}(a),(b) show the dispersion of the two lowest energy bands at two different chemical potentials ($\mu$) and
Figs.~\ref{Fig-theo}(c),(d) show the corresponding Fermi contours. For small values of $\mu$, only the lowest
band is occupied. Close to the $\Gamma$ point, it mostly consists of the a$_{\text{1g}}$ orbital. Away from
$\Gamma$, the orbital character changes and becomes anisotropic. At larger $\mu$, the second band is also
populated [Figs.~\ref{Fig-theo}(b),(d)]. For the range of $\mu$ relevant here, this band consists primarily
of one of the e$_{\text{g}}'$ orbitals and remains almost parabolic. Crucially, Fig.~\ref{Fig-theo}(e) shows
the monotonic variation of carrier density of the two bands as a function of $\mu$. The monotonic rise of the
second band population agrees quite well with the SdH data [Fig.~\ref{fig-expt}(a)], and supports our assumption 
that only this band gives rise to visible quantum oscillations, due to its higher mobility. 

\begin{figure}[t]
	\includegraphics[width=0.97\columnwidth]{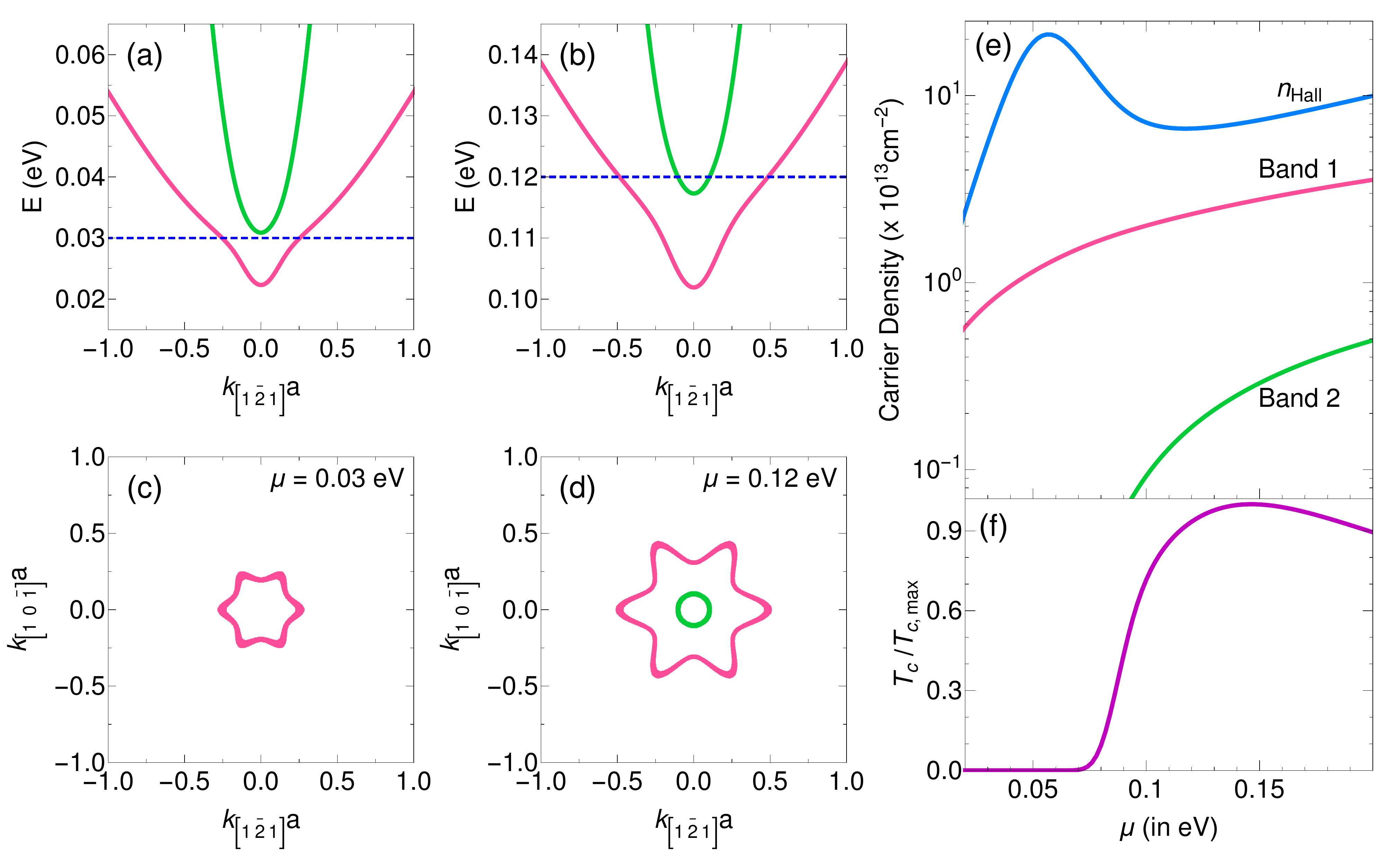}
	\caption{ Results of the theoretical model for the $(111)$ interface.
		(a),(b) Band structure for two different chemical potentials $\mu$ before
		($\mu$ = 0.03 eV) and after ($\mu$ = 0.12 eV) the second band starts
		getting occupied. The pink (green) line corresponds to band 1 (2), which
		is composed of a$_{\text{1g}}$ (e$_{\text{g}}^\prime$) orbitals close the $\Gamma$ point.
		The dashed blue line marks the Fermi energy.
		(c),(d) show the Fermi contours corresponding to the band structures in (a),(b)
		respectively. The outer surface (corresponding to band 1) is highly
		anisotropic at all $\mu$. (e) The carrier and Hall densities as a function of $\mu$.
		The carrier density is monotonic for both bands, while the Hall density is non-monotonic.
		The peak in $n_{\text{Hall}}$ occurs before the second band starts getting
		populated. This is in accordance with the experiment (Fig.~\ref{fig-expt})
		and indicates that it is due to the anisotropic shape of the lowest band.
		(f) Normalized superconducting critical temperature ($T_{\text{c}} / T_{\text{c,max}}$)
		as a function of $\mu$ within the single-band BCS model. }
	\label{Fig-theo}
\end{figure}
\begin{figure}[t]
	\includegraphics[width=0.97\columnwidth]{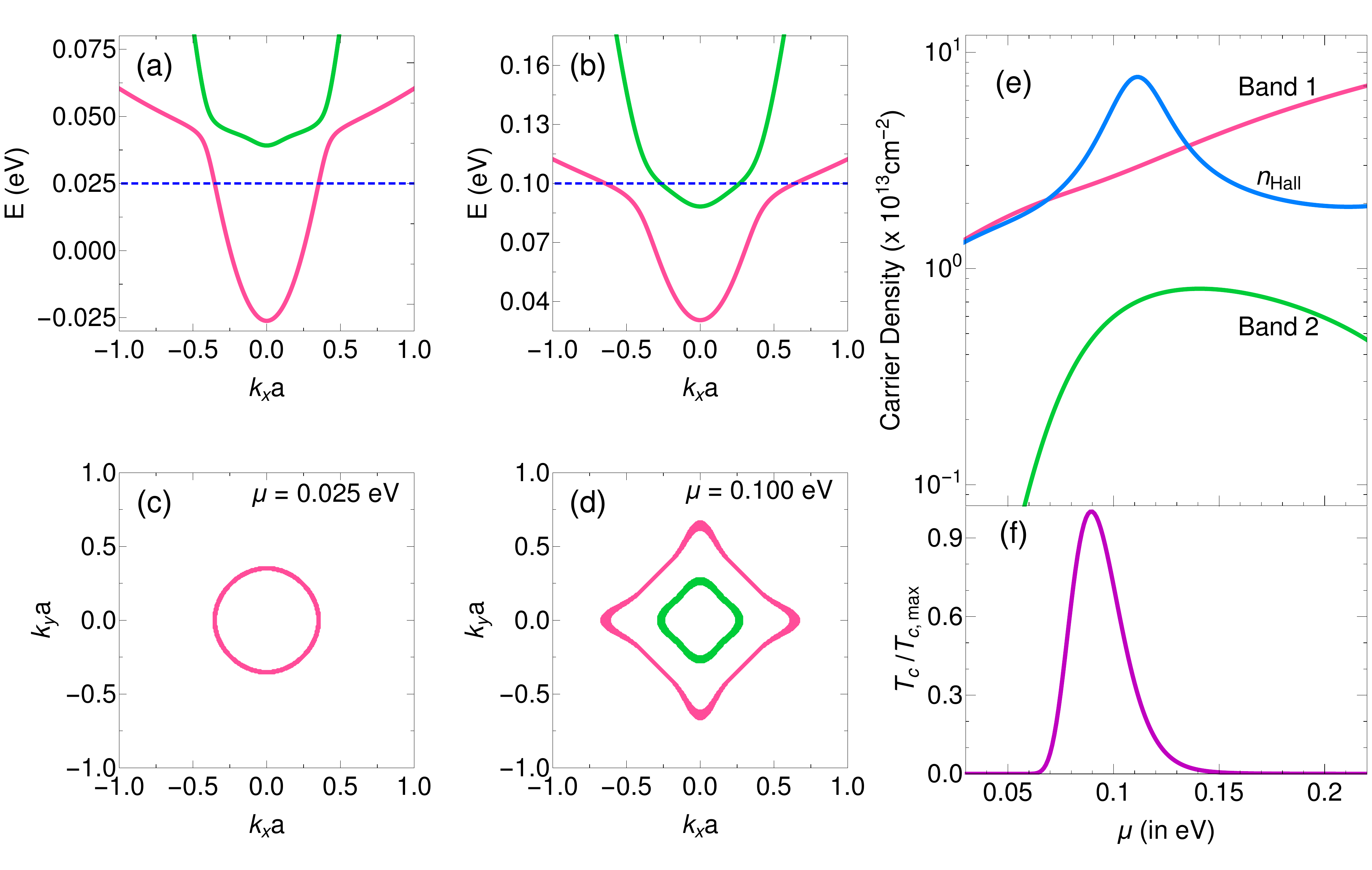}
	\caption{ Results of the theoretical model at $(001)$ interface.
		(a),(b) Bandstructure for two different chemical potentials $\mu$ before
		($\mu$ = 0.025 eV) and after ($\mu$ = 0.10 eV) the second band starts getting occupied.
		The pink (green) line corresponds to band 1 (2), which is composed of $xy$ (other t$_{2g}$) orbitals
		close to the $\Gamma$ point. The dashed blue
		line marks the Fermi energy. (c),(d) show the Fermi contours corresponding to
		the band structures in (a),(b) respectively. The outer surface (corresponding to band 1) is isotropic at low
		values of $\mu$. When the second band gets populated, the orbital character of the bands
		switches and the outer band becomes anisotropic. (e) The carrier and Hall densities
		as a function of $\mu$. The carrier density for second band is non-monotonic due to
		correlation-induced transfer of electrons from the lighter to heavier band. The Hall
		density is non-monotonic due to the anisotropy in the Fermi contour of the bands.
		(f) Normalized superconducting critical temperature ($T_{\text{c}} / T_{\text{c,max}}$)
		as a function of $\mu$  within the single-band BCS model. }
	\label{Fig-theo-copy}
\end{figure}

Upon increasing gate voltage the measured $n_{\text{Hall}}$ [Fig.~\ref{fig-expt}(a)] has a peak before the quantum oscillations
are visible. This means that this observed non-monotonicity must arise from the lowest band by itself. This is an important difference
between the $(001)$ and the $(111)$ interfaces that can be identified here because of our combination of
SdH and Hall measurements. Figs.~\ref{Fig-theo}(c),(d) show that the first band is non-parabolic and consists of
regions with both positive and negative curvature, throughout the range of relevant chemical potential.
This implies that a wavepacket gliding around the constant energy surface will give both electron-like
and hole-like contributions to the Hall conductivity. This is further complicated by the momentum-dependent
orbital character of the band at large filling. Under these conditions, the standard Drude relation
between inverse Hall coefficient and the carrier density of a single band
$[n_{\text{b}} = (eR_{\text{H}})^{-1}]$ is no longer valid and $n_{\text{Hall}}$
can differ significantly from the actual band population. Similarly the two band model, often used to fit 
Hall data for oxide interfaces, is valid only in case of two isotropic bands with no orbital mixing 
and therefore is not directly applicable for $(111)$ \LAO/\STO. 

To properly account for these features we compute the longitudinal and Hall conductivity
($\sigma_{\text{L}}$ and $\sigma_{\text{H}}$) using general expressions derived from the Boltzmann equation
assuming momentum dependent scattering times \cite{supp,Ong,hurd}. Specifically, we fix the orbital
lifetimes ($\tau_{\text{a}_{\text{1g}}}$ and $\tau_{\text{e}_{\text{g}\pm}^\prime}$) and assume the scattering
time for $m$th band to be, $\tau_m (\vec{K}) = \sum_{\sigma} \tau_{\text{a}_{\text{1g}}}
|\psi_m(\text{a}_{\text{1g}},\sigma,\vec{K})|^2 + \tau_{\text{e}_{\text{g}}'} \big(|\psi_m(\text{e}_{\text{g}+}^\prime,
\sigma,\vec{K})|^2 + |\psi_m(\text{e}_{\text{g}-}^\prime,\sigma,\vec{K})|^2 \big)$, where
$\psi_m$ is the self-consistent wavefunction for the $m$th band.
This allows $\tau_m(\vec{K})$ to trace the changes in orbital character along the Fermi contours.
Here we choose $\tau_{\text{e}_{\text{g}}^\prime} = 10\tau_{\text{a}_{\text{1g}}}$, so that the second band is more mobile.
While $\sigma_{\text{L}}$ and $\sigma_{\text{H}}$ depend on the orbital lifetimes
separately, and are thus harder to constrain by experimental data, 
$n_{\text{Hall}}$ ($\approx \sigma_{\text{L}}^2/\sigma_{\text{H}}$) depends only
on the ratio of the lifetimes. Therefore we show the variation of $n_{\text{Hall}}$ as a function
of $\mu$ in Fig.~\ref{Fig-theo}(e). The decent agreement of this theoretical result with experimental data
from Fig.~\ref{fig-expt}(a) implies that the experimental observations are a consequence of the shape and orbital
character of the lowest band.

We note that the Fermi contours in Fig.~\ref{Fig-theo}(c),(d) are similar to those reported 
for the $(111)$ surface of \STO~\cite{santander2014}. However, unlike the band structure in 
Fig.~\ref{Fig-theo}(a),(b) Ref.~\cite{santander2014} did not observe any splitting between the two lowest bands.  
This difference stems from the change in order of a$_{\text{1g}}$ and e$_{\text{g}}^{\prime}$ bands 
between the \STO~surface and \LAO/\STO~interface \cite{111DFT1,111DFT2,xld111}. In our model, this order 
is fixed by the sign of $\Delta_{\text{cf}}$, which we choose in accordance with
Ref.~\cite{xld111}. Using the opposite sign would provide a band structure similar to the one reported
in Ref.~\cite{santander2014}. 

Fig.~\ref{Fig-theo-copy} shows the results of a similar calculation for the
$(001)$ interface using a closely related model \cite{maniv100,MRS100}. Figs.~\ref{Fig-theo}(e)
and~\ref{Fig-theo-copy}(e) markedly differ in the behavior of the carrier density of band 2 at
the two interfaces: Here the population of band 2 ($n_\mathrm{SdH}$) is nonmonotonic, and the Hall 
density follows it [as opposed to monotonic SdH and maximal Hall number when band 2 is empty in $(111)$].
We stress that the nonmonotonic behavior of band 2 population at the $(001)$ interface
[Fig.~\ref{Fig-theo-copy}(e)] is not due to larger interaction terms
(the three largest parameters, $U$, $t$, and $t^\prime$, were taken to be equal in both cases).
Rather it occurs because at the $(001)$ interface the bands retain their original orbital
characters ($xy$, $yz$ and $xz$), which have a large difference in effective mass in the interface plane.
This generates a correlation-induced population transfer among the bands, because the
total energy can be minimized by transferring electrons from the lighter to heavier band \cite{maniv100,MRS100,smink100}. 
Since band 2 changes from heavy to light with increasing $\mu$ [Fig.~\ref{Fig-theo-copy}(a),(b)], it is first populated then depopulated.
In contrast, at the $(111)$ interface, all three t$_{\text{2g}}$ orbitals contribute equally to both
bands, and thus the effective band masses are not different enough for correlations
to induce population transfer. The nonmonotonic $n_\mathrm{Hall}$ in $(111)$ is rather the result of the greater Fermi 
contour curvature induced by the triangular symmetry, as compared to the square symmetry at $(001)$.

Finally, while our model does not account for the origin of superconductivity, we attempt to estimate the
superconducting critical temperature for our band structure using the single-band BCS
expression, $T_{\text{c}} = 1.13 T_{\theta} \exp(-\frac{1}{\rho_2 V_{\text{BCS}}})$ \cite{tinkham}.
Here we assume that the mobile band 2 has a higher contribution to the superconductivity, and
therefore use its density of states ($\rho_2$). $T_{\theta}$ is the Debye temperature of \STO~\cite{tdebye}, 
and $V_\mathrm{BCS}$ is set so that $T_{\text{c}}$ matches the experimental value at the maximum. Figs.~\ref{Fig-theo}(f) and
\ref{Fig-theo-copy}(f) show that we get good fits for the relative positions 
peaks in $T_{\text{c}}$ and $n_{\text{Hall}}$ with this simplistic model.

{\it Conclusions.---} We measured the variation of quantum oscillations frequency, Hall signal, and 
superconducting $T_{\text{c}}$ with gate voltage in $(111)$ \LAO/\STO~and
found it to be qualitatively different from the $(001)$ interface. Employing a tight-binding model with
on-site correlations, we calculated the band structure at both interfaces and showed that the
difference in the crystal structure leads to bands with different orbital character. In $(001)$ interface 
correlation-induced population transfer is the primary mechanism for the nonmonotonicity, while in $(111)$ 
it is the shape of the symmetry-induced Fermi contours.
This sets the stage for future investigation of the effect of this peculiar band structure on the 
superconductivity, magnetism, and ferroelectricity in these and related interfaces.

\begin{acknowledgments}
  U.K. and P.K.R. contributed equally to this work. U.K. was supported by the Raymond and
  Beverly Sackler Faculty of Exact Sciences at Tel Aviv University and the Raymond and Beverly Sackler 
  Center for Computational Molecular and Material Science. Y.D. and M.G. were
  supported by the ISF (Grants No. 382/17 and 227/15), BSF (Grants No. 2014047 and 2016224),
  GIF (Grant No. I-1259-303.10) and the Israel Ministry of Science and Technology 
  (Contract No. 3-12419). Part of this work was done at the High-Field Magnet Laboratory 
  (HFML-RU/NWO), member of the European Magnetic Field Laboratory (EMFL).
\end{acknowledgments}

\onecolumngrid
\clearpage

\setcounter{affil}{0}
\renewcommand{\thefigure}{S\arabic{figure}}
\setcounter{figure}{0}
\renewcommand{\theequation}{S\arabic{equation}}
\setcounter{equation}{0}
\renewcommand\thesection{S\arabic{section}}
\setcounter{section}{0}

\title{Supplemental material for ``Symmetry and correlation effects on band structure explain the anomalous transport properties of $(111)$ LaAlO$_3$/SrTiO$_3$''}

\begin{abstract}
This set of supplemental materials provides additional details about our theoretical 
model and the analysis of transport data. Section I describes the constraints on the
one-body matrix elements due to the crystal symmetries of the $(111)$ interface.
Section II gives a detailed description of the interface model and the calculation of 
the conductance, along with results regarding the effect of accounting for additional 
layers. Finally, in Section III we present additional magnetotransport data along 
with the analysis.
\end{abstract}

\maketitle

\section{I.\,\,\,\,\,\,      Structure and Symmetry of the $(111)$ interface}

\begin{figure}[t]
\includegraphics[width=0.7\columnwidth]{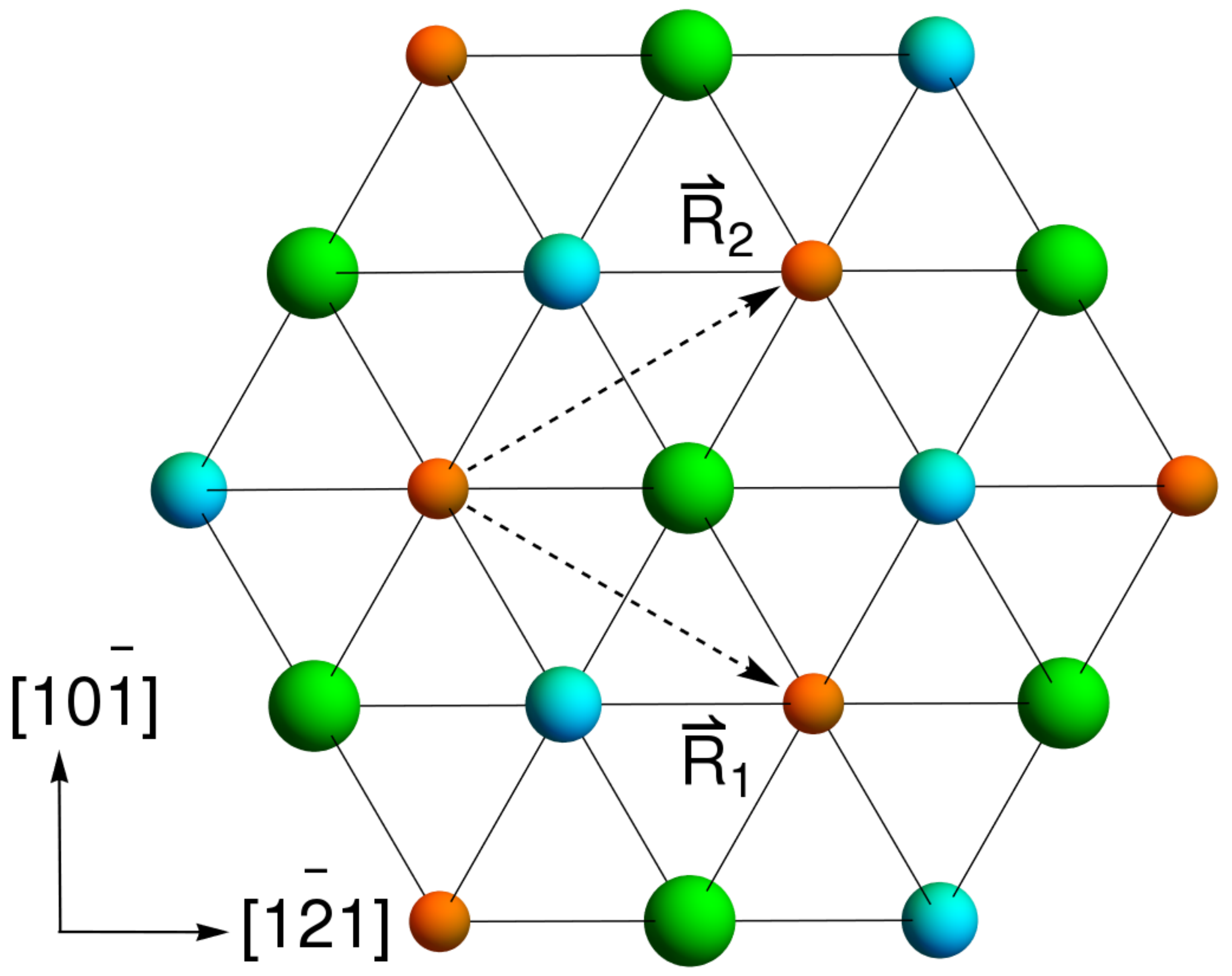} 
  \caption{ Projection of a cubic lattice onto the plane normal to the $(111)$ direction. 
  A unit cell with 3 inequivalent sites (denoted by green $\equiv \text{Ti}_1$, orange 
  $\equiv \text{Ti}_2$ and blue $\equiv \text{Ti}_3$) forms a triangular 
  lattice stacked along the $\hat{Z}$ direction. $\vec{R}_{\text{1}}$ and $\vec{R}_{\text{2}}$ are the new 
  (in-plane) lattice vectors for the triangular lattice. }
\label{Fig-structure}
\end{figure}

As described in the main text, the low-energy conduction bands in bulk \STO~are
composed of the t$_{\text{2g}}$ orbitals of Ti, which form a cubic structure
(neglecting the structure distortions at low temperatures). Fig.~\ref{Fig-structure} 
shows that the projection of a cube into a plane normal to the $(111)$ direction is 
a stack of triangular lattices with 3 inequivalent sites (labelled as Ti$_{i}$). 
The new (in-plane) lattice vectors are
\begin{align}   
  \vec{R}_{\text{1,2}} &= \sqrt{2} a \left(\frac{\sqrt{3}}{2}\hat{X} \mp \frac{\hat{Y}}{2} \right),
\end{align}
where $a$ is the lattice constant (of the cubic lattice) and $\hat{X},\hat{Y}$ 
are unit vectors along the $[1\bar{2}1],[10\bar{1}]$ directions, respectively. We also define
$\hat{Z}$ as the $[111]$ direction. The change in crystal structure (from cubic to triangular) introduces 
a new (in-plane) crystal field at the interface, which lifts the degeneracy of the t$_{\text{2g}}$ 
orbitals and mixes them to form, 
\begin{align}
  |\text{a}_{\text{1g}} \rangle &= \frac{1}{\sqrt{3}} \big[ |xy\rangle + |yz\rangle + |xz\rangle \big], \\
  |\text{e}_{\text{g}+}^{\prime}\rangle &= \frac{1}{\sqrt{3}} \big[ |xy\rangle + \omega |yz\rangle + \omega^2 |xz\rangle \big], \\
  |\text{e}_{\text{g}-}^{\prime}\rangle &= \frac{1}{\sqrt{3}} \big[ |xy\rangle + \omega^{-1} |yz\rangle + \omega^{-2} |xz\rangle \big], 
\end{align}

\noindent
where $\omega = e^{i \frac{2\pi}{3}}$ is the cubic root of unity. Below we describe how these
orbitals behave under the symmetry transformations relevant to this system.

\subsubsection*{Spatial Symmetries}
The triangular lattice formed by $\big(\vec{R}_1,\vec{R}_2\big)$ is invariant under the 
C$_{3v}$ group, which can be generated by two operators: 

\noindent
  1. {\it Rotation about $\hat{Z}$ by $\frac{2\pi}{3}$} : 
  The $|\text{a}_{\text{1g}}\rangle$ orbital is invariant under this transformation, while the others pick up
  a phase,
  \begin{align} |\text{e}_{\text{g}\pm}^{\prime}\rangle \rightarrow \omega^{\pm 2} |\text{e}_{\text{g}\pm}^{\prime}\rangle. \end{align}
  Similarly, choosing $\hat{Z}$ as the spin quantization axis, the spin states transform as,
  \begin{align} \{|\ua\rangle, |\da\rangle\} \rightarrow \{-\omega |\ua\rangle, -\omega^{-1} |\da\rangle\}. \end{align}

\noindent
  2. {\it Reflection about $\hat{X}-\hat{Z}$ plane} : Reflection can be thought of as 
  inversion about the origin followed by a rotation of $\pi$ about $\hat{Y}$. Again, 
  $|\text{a}_{\text{1g}}\rangle$ is invariant under such a reflection, while the other states transform into
  each other, 
  \begin{align} |\text{e}_{\text{g}\pm}^{\prime}\rangle \rightarrow \omega^{\pm 1} |\text{e}_{\text{g}\mp}^{\prime}\rangle. \end{align}
  Similarly, the spin states are also exchanged,
  \begin{align} \{|\ua\rangle, |\da\rangle\} \rightarrow \{\omega^{-1} |\da\rangle, -\omega |\ua\rangle\}. \end{align}

\subsubsection*{Time-Reversal}
The time-reversal operator involves complex conjugation followed by a rotation of the spins 
by $\pi$ along some axis. We choose to rotate along $-\frac{\sqrt{3}}{2}\hat{X} - \frac{\hat{Y}}{2}$
so that the spin states transform (under time-reversal) as, 
\begin{align} \{|\ua\rangle, |\da\rangle\} \rightarrow \{ |\da\rangle, -|\ua\rangle\}. \end{align}
The orbital states are eigenstates of $\hat{L}_{Z}$ (angular momentum along $\hat{Z}$) and 
hence also transform under time-reversal. The a$_{\text{1g}}$ corresponds to $m=0$ and is therefore invariant, 
while the others (corresponding to $m = \pm 1$) transform as,
\begin{align} |\text{e}_{\text{g}\pm}^{\prime}\rangle \rightarrow  |\text{e}_{\text{g}\mp}^{\prime}\rangle. \end{align}

\subsubsection*{Constraints on one-body matrix elements}
In this work, we assume that the final ground state of the system is invariant under translations 
(by $\vec{R}_{1,2}$), time-reversal and all the spatial symmetries of the crystal structure ($C_{3v}$). 
This invariance introduces some constraints on the (on-site) one-body matrix elements, 
\begin{equation}
  V_{\sigma \sigma^{\prime}}^{m m^{\prime}} (\vec{R},i) = \langle c^{\dagger}_{i m \sigma} 
  (\vec{R}) c_{i m^{\prime} \sigma^{\prime}} (\vec{R}) \rangle,
\end{equation}
where $\vec{R}$ is the position of the unit cell, $i$ labels the different atoms (Ti$_i$) of the unit cell, 
$m,m^{\prime}$ are the on-site orbitals ($\text{a}_{\text{1g}},\text{e}_{\text{g}\pm}^{\prime}$) and 
$\sigma,\sigma^{\prime}$ are the spin states. We use these constraints to simplify the mean-field ansatz. 
Since the two-body term involved in our calculation is an on-site term, we do not need the constraints on
other matrix elements. 

Due to translation symmetry, the matrix elements are independent of $\vec{R}$ (but not of $i$). 
We note that the transformations considered here only mix states on a given site ($i$) of the unit 
cell, i.e., $i$ does not change under the symmetry operations described below. 

1. {\it Time-Reversal} : For a system invariant under time-reversal ($\mathcal{T}$), the matrix elements 
must satisfy,
\begin{equation}
  \langle \psi | \phi \rangle = \langle \tilde{\phi} | \tilde{\psi} \rangle = \langle \tilde{\psi} | \tilde{\phi} \rangle^{*},
\end{equation}
where $|\tilde{\psi}\rangle = \mathcal{T} |\psi\rangle$ and $|\tilde{\phi}\rangle = \mathcal{T} |\phi\rangle$. 
This implies that due to time-reversal symmetry, 
\begin{equation} 
  V_{\sigma \sigma^{\prime}}^{m m^{\prime}} (i) = \pm V_{\tilde{\sigma}^{\prime} \tilde{\sigma}}^{\tilde{m}^{\prime} \tilde{m}} (i),
\end{equation}
where the $+$ ($-$) sign appears if $\sigma = \sigma^{\prime}$ (otherwise), $\tilde{\sigma}$ is the spin state
opposite to $\sigma$ and $\tilde{m}$ is the orbital state related to $m$ as described in equation (S10).
This means that the occupancies of several levels are related to each other,
\begin{align}
  V_{\ua \ua}^{\text{a}_{\text{1g}} \text{a}_{\text{1g}}} (i) &= 
  V_{\da \da}^{\text{a}_{\text{1g}} \text{a}_{\text{1g}}} (i) = N_{\text{a}_{\text{1g}}} (i), \\
  V_{\ua \ua}^{\text{e}_{\text{g}+}^{\prime} \text{e}_{\text{g}+}^{\prime}} (i) &= 
  V_{\da \da}^{\text{e}_{\text{g}-}^{\prime} \text{e}_{\text{g}-}^{\prime}} (i) = N_{\text{e}_{\text{g}+}^{\prime}} (i), \\ 
  V_{\ua \ua}^{\text{e}_{\text{g}-}^{\prime} \text{e}_{\text{g}-}^{\prime}} (i) &= 
  V_{\da \da}^{\text{e}_{\text{g}+}^{\prime} \text{e}_{\text{g}+}^{\prime}} (i) = N_{\text{e}_{\text{g}-}^{\prime}} (i). 
\end{align}
Additionally, the following matrix elements must necessarily vanish (for all $i$),
\begin{align}
  V_{\ua \da}^{\text{a}_{\text{1g}} \text{a}_{\text{1g}}} = \, &0 = 
  V_{\da \ua}^{\text{a}_{\text{1g}} \text{a}_{\text{1g}}},\\
  V_{\ua \da}^{\text{e}_{\text{g}+}^{\prime} \text{e}_{\text{g}-}^{\prime}} = \, &0 = 
  V_{\da \ua}^{\text{e}_{\text{g}+}^{\prime} \text{e}_{\text{g}-}^{\prime}}, \\
  V_{\ua \da}^{\text{e}_{\text{g}-}^{\prime} \text{e}_{\text{g}+}^{\prime}} = \, &0 = 
  V_{\da \ua}^{\text{e}_{\text{g}-}^{\prime} \text{e}_{\text{g}+}^{\prime}}. 
\end{align}

2. {\it Rotations} : Using equations (S5) and (S6) we find that invariance under rotations forces 
yet more matrix elements to vanish (at all $i$),
\begin{align}
  V_{\ua \da}^{\text{e}_{\text{g} s}^{\prime} \text{e}_{\text{g} s}^{\prime}} = \, &0 = 
  V_{\da \ua}^{\text{e}_{\text{g} s}^{\prime} \text{e}_{\text{g} s}^{\prime}}, \\ 
  V_{\sigma \sigma}^{\text{e}_{\text{g}+}^{\prime} \text{e}_{\text{g}-}^{\prime}} = \, 
  &0 = V_{\sigma \sigma}^{\text{e}_{\text{g}-}^{\prime} \text{e}_{\text{g}+}^{\prime}}, \\
  V_{\sigma \sigma}^{\text{a}_{\text{1g}} \text{e}_{\text{g} s}^{\prime}} = \, &0 
  = V_{\sigma \sigma}^{\text{e}_{\text{g} s}^{\prime} \text{a}_{\text{1g}}}, \\
  V_{\ua \da}^{\text{a}_{\text{1g}} \text{e}_{\text{g}-}^{\prime}} = \, &0 
  = V_{\ua \da}^{\text{e}_{\text{g}+}^{\prime} \text{a}_{\text{1g}}}, \\
  V_{\da \ua}^{\text{a}_{\text{1g}} \text{e}_{\text{g}+}^{\prime}} = \, &0 
  = V_{\da \ua}^{\text{e}_{\text{g}-}^{\prime} \text{a}_{\text{1g}}}. 
\end{align}
for all $s = \pm$ and $\sigma = \ua,\da$. 

3. {\it Reflections} : The only non-zero matrix elements with different spin states are related 
to each other through the following sequence of operations (again at all $i$),
\begin{align}
  V_{\ua \da}^{\text{a}_{\text{1g}} \text{e}_{\text{g}+}^{\prime}} &= 
  - V_{\ua \da}^{\text{e}_{\text{g}-}^{\prime} \text{a}_{\text{1g}}}  
    \,\, \text{ by time-reversal invariance} \nonumber \\
    &= - \big( V_{\da \ua}^{\text{a}_{\text{1g}} \text{e}_{\text{g}-}^{\prime}} \big)^{*}  
    \,\, \text{ by hermitian-conjugation} \nonumber \\
    &= - \big( - V_{\ua \da}^{\text{a}_{\text{1g}} \text{e}_{\text{g}+}^{\prime}} \big)^{*}  
    \,\, \text{ due to reflection invariance} \nonumber \\
    &= \big( V_{\ua \da}^{\text{a}_{\text{1g}} \text{e}_{\text{g}+}^{\prime}} \big)^{*}.
\end{align}
The last line implies that these matrix elements are all real. Then we define the only spin-mixing
average allowed by symmetry as
\begin{align}
  N_{\text{SO}} (i) = V_{\ua \da}^{\text{a}_{\text{1g}} \text{e}_{\text{g}+}^{\prime}} (i) 
  = - V_{\da \ua}^{\text{a}_{\text{1g}} \text{e}_{\text{g}-}^{\prime}} (i).
\end{align}

Therefore, the crystal structure at the $(111)$ interface allows only four (real and independent) 
one-body matrix elements (per layer) :
$N_{\text{a}_{\text{1g}}} (i)$, $N_{\text{e}_{\text{g}+}^{\prime}} (i)$, $N_{\text{e}_{\text{g}-}^{\prime}} (i)$, and  
$N_{\text{SO}} (i)$.

\section{II.\,\,\,\,\,\,      Hamiltonian and conductivity at the $(111)$ interface}

As described above, the $(111)$ interface of \STO/\LAO~has a triangular crystal structure with 
a complex unit cell (cf. Fig. S1). In order to correctly represent the connectivity of the different Ti atoms
(in the underlying cubic lattice), we first write a Hamiltonian for the bulk and then adapt it to 
describe the interface. 

\subsubsection*{Bulk Hamiltonian}

In the bulk, we employ a tight-binding Hamiltonian based on a model of Refs.~\cite{MacDonald-11,MacDonald-12}. 
It can be written as 
$H_{\text{b}}(\vk) = H_{\text{0}}(\vk) + H_{\text{SO}}$, where $H_{\text{SO}} = \Delta_{\text{SO}} 
\vec{L} \cdot \vec{s}$ is the atomic spin-orbit (SO) coupling and $H_{\text{0}}(\vk)$ is the 
kinetic term composed of nearest-neighbor (NN) hopping. 

We express $H_{\text{b}}$ in terms of the a$_{\text{1g}}$ and e$_{\text{g}\pm}^{\prime}$
orbitals and use $\hat{Z}$ as the spin-quantization axis on the trilayer triangular lattice (Fig.~S1),
which is equivalent to the cubic lattice. 
The three lattice vectors for this structure are $\vec{R}_{1,2}$ defined in Eq.~(S1) and 
$\vec{R}_3 = \sqrt{3} a \hat{Z}$. The three-dimensional Brillouin zone is defined by the reciprocal 
lattice vectors,
$\vec{G}_{1,2} = \frac{\sqrt{2}}{\sqrt{3} a} \left(\frac{\hat{K}_{X}}{2} \mp 
\frac{\sqrt{3}}{2}\hat{K}_{Y}\right)$
and $\vec{G}_3 = \frac{1}{\sqrt{3} a} \hat{K}_{Z}$.
To avoid confusion, we denote the three-dimensional (bulk) momentum by $\vk = (k_1,k_2,k_3)$ and the 
two-dimensional surface momentum by $\vK = (K_1,K_2)$, with $k_{a}, K_{a} \in [-\pi,\pi]$. Labelling
the three layers as $i = 1,2,3$, 
we can write the NN term as an 18$\times$18 matrix in the basis, $\{ |a_{\text{1g}}\rangle, 
|e^{\prime}_{\text{g}+}\rangle, |e^{\prime}_{\text{g}-}\rangle \} \otimes \{ |\text{Ti}_{1}\rangle, |\text{Ti}_{2}\rangle, 
|\text{Ti}_{3}\rangle \} \otimes \{|\ua\rangle, |\da\rangle \}$ as
\begin{align} 
  &H_{\text{0}}(\vk) =  \left( \begin{array}{ccc} 
    \tilde{A}_{2}(\vk) & \tilde{B}_{2}(\vk) & \tilde{B}_{2}^{\dagger}(\vk) \\
   \tilde{B}_{2}^{\dagger}(\vk) & \tilde{A}_{2}(\vk) & \tilde{B}_{2}(\vk) \\
    \tilde{B}_{2}(\vk) & \tilde{B}_{2}^{\dagger}(\vk) & \tilde{A}_{2}(\vk) 
  \end{array} \right) \otimes \mathcal{I}_2 ,
\end{align}
where the block matrices $\tilde{A}_2(\vk)$ and $\tilde{B}_2(\vk)$ are, 
\begin{widetext}
\begin{align}
  \tilde{A}_2(\vk) &= -\frac{(2t+t^{\prime})}{3} \left( \begin{array}{ccc} 
      0 & e^{-i k_2} f_{0}(\vk) & e^{-i k_3} f_{0}(-\vk) \\
      e^{i k_2} f_{0}(-\vk) & 0 & e^{-i k_1} f_{0}(\vk) \\
  e^{i k_3} f_{0}(\vk) & e^{i k_1} f_{0}(-\vk) & 0 \end{array} \right), \\
  \tilde{B}_2(\vk) &= \omega^2 \frac{(t-t^{\prime})}{3} \left( \begin{array}{ccc} 
      0 & e^{-i k_2} f_{\omega}(\vk) & e^{-i k_3} f_{\omega}(-\vk) \\
      e^{i k_2} f_{\omega}(-\vk) & 0 & e^{-i k_1} f_{\omega}(\vk) \\
      e^{i k_3} f_{\omega}(\vk) & e^{i k_1} f_{\omega}(-\vk) & 0 \end{array} \right),
\end{align}
\end{widetext}
where $t$ and $t^{\prime}$ are the light and heavy NN hopping amplitudes, 
$f_{0}(\vk) = 1 + e^{i k_1} + e^{i k_2}$ and $f_{\omega}(\vk) = 1 + \omega e^{i k_1} + \omega^2 e^{i k_2}$.
The terms with $e^{i k_3}$ represent the NN terms connecting different trilayers, while the
others represent kinetic hopping within the same trilayer.

While bulk \STO~is well described by $H_{\text{b}}$, previous works \cite{joshua2012,ruhman2014,maniv100,MRS100,
Satoshi11,Satoshi13,Satoshi18} have found that the next-nearest-neighbor (NNN) hopping terms play an 
important role at the interface. The multi-orbital electronic structure of \STO~can give rise to many possible 
NNN terms. Here we use the one which is expected to be the largest \cite{Satoshi11}. Using the
same basis as for $H_{\text{0}}$,
\begin{align} \label{eq-NNN}
  &H_{\text{NNN}}(\vk) =  \left( \begin{array}{ccc} 
    \tilde{A}_{3}(\vk) & \tilde{B}_{3}(\vk) & \tilde{B}_{3}^{\dagger}(\vk) \\
   \tilde{B}_{3}^{\dagger}(\vk) & \tilde{A}_{3}(\vk) & \tilde{B}_{3}(\vk) \\
    \tilde{B}_{3}(\vk) & \tilde{B}_{3}^{\dagger}(\vk) & \tilde{A}_{3}(\vk) 
  \end{array} \right) \otimes \mathcal{I}_2,
\end{align}
where the block matrices $\tilde{A}_3(\vk)$ and $\tilde{B}_3(\vk)$ are, 
\begin{widetext}
\begin{align}
  \tilde{A}_3(\vk) &= -\frac{t^{\prime\prime}}{3} \left( \begin{array}{ccc} 
    2 \epsilon_{0}(\vk) & e^{-i (k_3 + k_2)} f_{0}(\vk) & f_{0}(-\vk) \\
      e^{i (k_3 + k_2)} f_{0}(-\vk) & 2 \epsilon_{0}(\vk) & e^{-i (k_3 + k_1)} f_{0}(\vk) \\
      f_{0}(\vk) & e^{i (k_3 + k_1)} f_{0}(-\vk) & 2 \epsilon_{0}(\vk) \end{array} \right), \\
  \tilde{B}_3(\vk) &= -\omega^2 \frac{t^{\prime\prime}}{3} \left( \begin{array}{ccc} 
      2 \epsilon_{\omega}(\vk) & e^{-i (k_3 + k_2)} f_{\omega}(\vk) & f_{\omega}(-\vk) \\
      e^{i (k_3 + k_2)} f_{\omega}(-\vk) & 2 \epsilon_{\omega}(\vk) & e^{-i (k_3 + k_1)} f_{\omega}(\vk) \\
      f_{\omega}(\vk) & e^{i (k_3 + k_1)} f_{\omega}(-\vk) & 2 \epsilon_{\omega}(\vk) \end{array} \right),
\end{align}
\end{widetext}
where $t^{\prime\prime}$ is the NNN hopping and $\epsilon_{0} (\vk) = \cos(k_1) + \cos(k_2) + \cos(k_1 - k_2)$ and 
$\epsilon_{\omega} (\vk) =  \cos(k_1 - k_2) + \omega \cos(k_2) + \omega^2 \cos(k_1) $. 
We note that the NNN term in Eq.~(\ref{eq-NNN}) is diagonal in the basis of $xy$, $yz$ and $xz$ 
orbitals. Therefore, at the $(001)$ interface it does not play an important role. In that case, a different
term which mixes the orbitals is more important \cite{joshua2012,ruhman2014,maniv100,MRS100} 
(even though it is smaller in amplitude).

Finally, the on-site atomic SO coupling is given by (in the basis $\{ |a_{\text{1g}}\rangle, 
|e^{\prime}_{\text{g}+}\rangle, |e^{\prime}_{\text{g}-}\rangle \} \otimes \{|\ua\rangle, |\da\rangle \}$),
\begin{align}
  &H_{\text{SO}} = \frac{\Delta_{\text{so}}}{2} \left( \begin{array}{ccc} 
    0 & -\sqrt{2}\sigma^{+} & \sqrt{2}\sigma^{-} \\
     -\sqrt{2}\sigma^{-} & - \sigma_{z} & 0 \\
     \sqrt{2}\sigma^{+} & 0 & \sigma_{z} 
  \end{array} \right), 
\end{align}
where $\sigma^{\pm} = \frac{1}{2} (\sigma_{x} \pm i \sigma_{y})$, and $\sigma_{x,y,z}$ are the 
Pauli matrices. 
The SO coupling splits the degeneracy of the t$_{\text{2g}}$ orbitals and forms two sets of states with 
spin $\frac{3}{2}$ and $\frac{1}{2}$. This is because the t$_{\text{2g}}$ orbitals form a $l=1$ multiplet when 
mixing with e$_{\text{g}}$ orbitals is ignored \cite{t2g}. The sign of spin-orbit coupling 
($\Delta_{\text{SO}} > 0$) is chosen such that in the bulk, the spin-$\frac{3}{2}$ multiplet is 
lower in energy and the spin-$\frac{1}{2}$ is higher in energy. This is in accordance with ab-initio 
studies on the bulk of \STO~\cite{spin-orbit}. 

\subsubsection*{Interface Hamiltonian}

\begin{figure}[b]
\includegraphics[width=0.99\columnwidth]{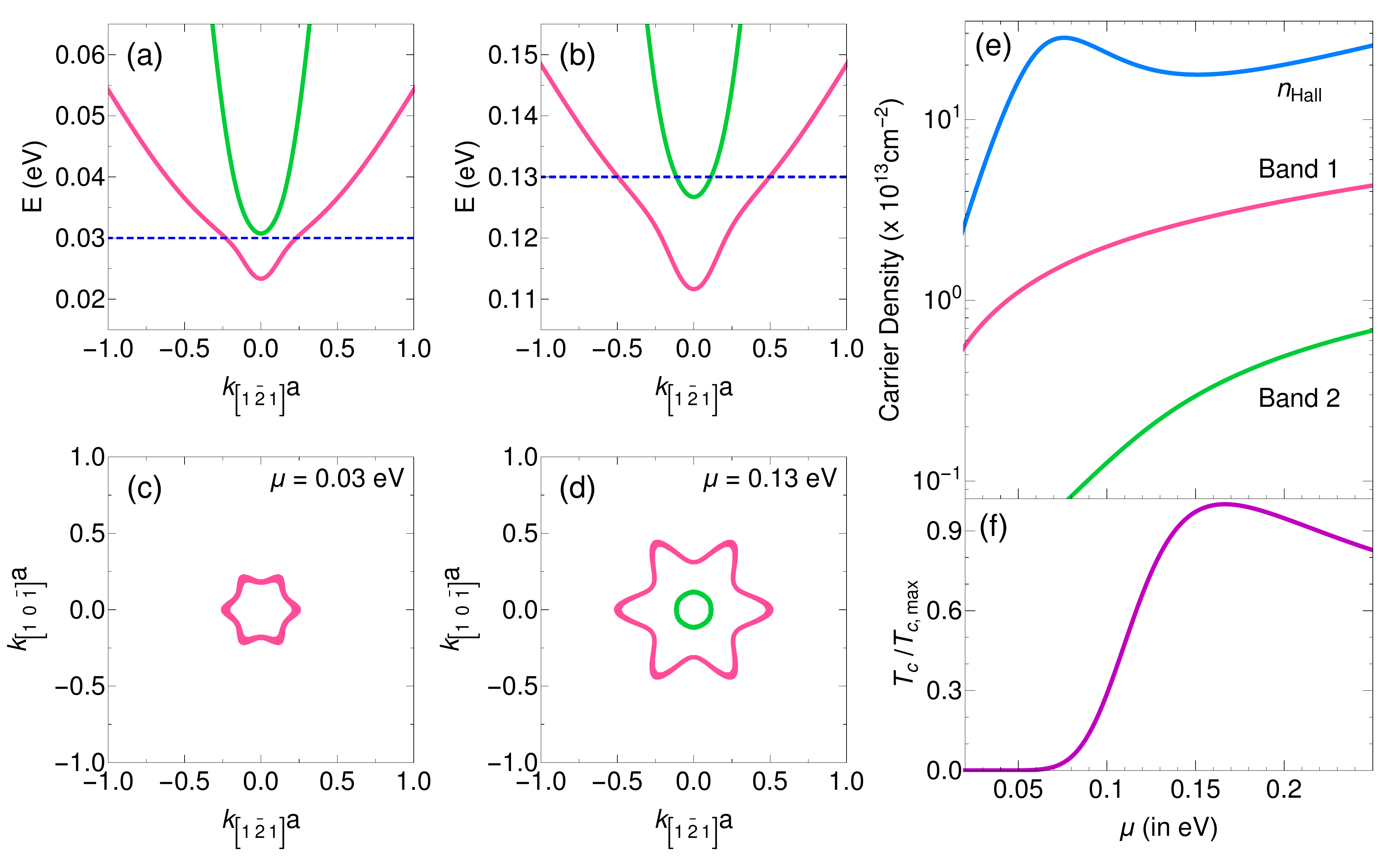} 
\caption{Band structure and transport coefficients at the $(111)$ \LAO/\STO~interface. Six layers
  were included in the calculation with a confining potential $V_{\text{c}} =$ 100 meV, while 
  other parameters were identical to those used for Fig. 3 of the main text. The changes in the 
  behavior are insignificant. }
\label{Fig-Supp-6L}
\end{figure}

At the $(111)$ interface we start with the bulk Hamiltonian $H_{\text{b}} + H_{\text{NNN}}$ defined above 
on a small number of layers along the $\hat{Z}$ direction. Now the system is periodic only in the $\hat{X}$-$\hat{Y}$ 
plane and the number of Ti atoms in a unit cell is the number of layers included in the calculation. As
described in the main text, we only keep three layers in this work. In this case, $H_{\text{0}} + H_{\text{NNN}}$ reduce 
to the matrices defined in equations (1)$-$(3) of the main text. Below we will show that incorporating more
layers does not modify our results in a significant way (cf. Fig.~\ref{Fig-Supp-6L}).

Now the spin-orbit term, being an on-site coupling, does not change at the interface.
The confining potential at the interface is of course different on each layer since they are separated 
in the $\hat{Z}$ direction. Here we model confinement as a linearly increasing potential and denote its difference
between two adjacent layers by $V_{\text{C}}$. In the basis $ \{ |\text{Ti}_{1}\rangle, |\text{Ti}_{2}\rangle, 
|\text{Ti}_{3}\rangle \} \otimes \{|\ua\rangle, |\da\rangle \}$ it can be written as, 
\begin{align}
  &H_{\text{V}} =  \left( \begin{array}{ccc} 
    2V_{\text{C}} & 0 & 0 \\
    0 & V_{\text{C}} & 0 \\
    0 & 0 & 0 
  \end{array} \right) \otimes \mathcal{I}_2 .
\end{align}

As described in the main text, the change of symmetry at the $(111)$ interface gives rise to a new
in-plane crystal field. In the basis $\{ |a_{\text{1g}}\rangle, 
|e^{\prime}_{\text{g}+}\rangle, |e^{\prime}_{\text{g}-}\rangle \} \otimes \{|\ua\rangle, |\da\rangle \}$ it is,
\begin{align}
  &H_{\text{cf}} =  \frac{\Delta_{\text{cf}}}{2} \left( \begin{array}{ccc} 
    -2 & 0 & 0 \\
    0 & 1 & 0 \\
    0 & 0 & 1 
  \end{array} \right) \otimes \mathcal{I}_2 .
\end{align}

Finally, we include correlation effects in the model by adding an on-site Hubbard interaction of the form,
\begin{align} 
  \sum_{\text{r}} \sum_{\text{I} \neq \text{J}} U n_{\text{rI}} n_{\text{rJ}} ,
\end{align}
where I,J denote both orbital and spin quantum numbers. For simplicity, we assume that the strength of 
intra-orbital and inter-orbital repulsion is equal. This reduces the number of free
parameters in the problem but (as we have verified) does not affect the results in any substantial way. We treat the two-body term in the 
Hartree-Fock approximation assuming that the ground-state is invariant under time-reversal and the spatial 
symmetries of the interface (while some of the spatial symmetries are broken at the relevant temperature, we have checked that
the corresponding modifications to the Hamiltonian and the results are rather small). 
As described in section I, under this ansatz there are only four independent real one-body
averages (for each layer). Defining
$N_{\pm}(i) = N_{\text{e}_{\text{g}+}^{\prime}}(i) \pm N_{\text{e}_{\text{g}-}^{\prime}}(i)$ and
$N_{\text{T}}(i) = 2\big(N_{\text{a}_{\text{1g}}}(i) + N_{+}(i) \big)$, 
the Hartree-Fock terms on $i^{\text{th}}$ layer are,
\begin{align}
  H_{\text{H}}(i) &= U \big[N_{\text{T}}(i)\, \mathcal{I}_6 + H_{\text{H1}}(i) + H_{\text{H2}}(i)\big], \\
  H_{\text{H1}}(i) &= -\frac{1}{2} \left( \begin{array}{ccc} 
    2N_{\text{a}_{\text{1g}}}(i) & 0 & 0 \\
    0 & N_{+}(i) & 0 \\
    0 & 0 & N_{+}(i) 
  \end{array} \right) \otimes \mathcal{I}_2 ,\\
  H_{\text{H2}}(i) &= -\frac{1}{2} N_{-}(i) 
  \left( \begin{array}{ccc} 
    0 & 0 & 0 \\
    0 & \sigma_z & 0 \\
    0 & 0 & -\sigma_z 
  \end{array} \right),\\
  H_{\text{F}}(i) &=  -U N_{\text{SO}}(i) \left( \begin{array}{ccc} 
    0 & \sigma^{+} & -\sigma^{-} \\
    \sigma^{-} & 0 & 0 \\
    -\sigma^{+} & 0 & 0 
  \end{array} \right).
\end{align}

\begin{figure*}
\includegraphics[width=0.85\textwidth]{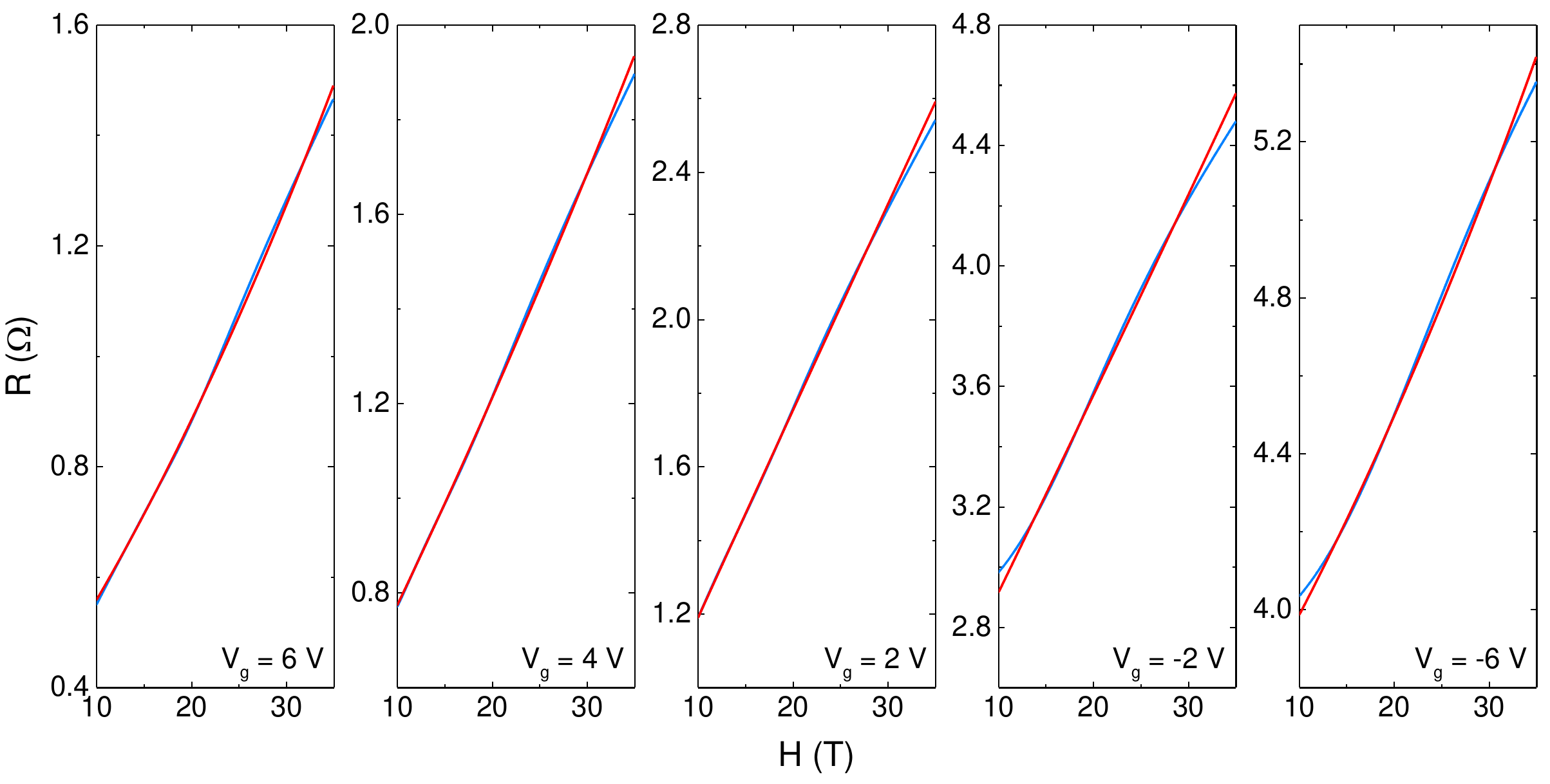} 
\caption{Perpendicular magnetoresistance (blue) at $T =$ 0.34 K for $V_{\text{g}}$ = 
  6, 4, 2, $-$2, and $-$6 V, together with fits to a quadratic ($a_0 + a_{\text{1}}H + a_{\text{2}}H^2$) 
  function (red).}
\label{Fig-Supp-RV}
\end{figure*}

\noindent
Clearly, the Hartree-Fock terms renormalize the spin-orbit and crystal-field couplings separately on each layer.
To simplify the calculations we further add an overall constant $\epsilon_{\mathcal{I}}$ to the Hamiltonian 
so that (at $U = 0$) the minimum of the lowest band is around zero. For three layers and $t^{\prime\prime} \ll t,V_c$, 
$\epsilon_{\mathcal{I}}$ is
\begin{align}
   \frac{1}{2}\big(4t^{\prime\prime} + \Delta_{\text{SO}}-\Delta_{\text{cf}}\big) 
  -V_c + \sqrt{V_c^2 + 2(2t+t^{\prime})^2}.
\end{align}

\subsubsection*{Transport Coefficients}

As explained in the main text, the band structure resulting from the self-consistent calculation of our model
is far from isotropic. The six-fold symmetric constant energy surfaces have regions with both positive and negative 
curvature (Fig. 3(a),(b) of the main text). Therefore in the semi-classical picture, a wave-packet gliding along the Fermi
contour (under the effect of a perpendicular magnetic field) would behave as an electron and as a hole at different 
momenta (different times). The often-employed 
Drude theory is not valid under these conditions and therefore we use more general expressions, derived from the
Boltzmann equation \cite{Ong,hurd}, for the longitudinal ($\sigma_{\text{L}}$) and Hall ($\sigma_{\text{H}}$) conductivity, 
\begin{widetext}
  \begin{align}
    \sigma_{L} &= -e^2 \sum_{m} \int_{BZ} \frac{d^2 K}{(2\pi)^2} \big(\partial_{\mu} f[\epsilon_m(\vK)]\big)
    \tau_m (\vK) \frac{1}{2}\big(\big[v^X_{m}(\vK)\big]^2 + \big[v^Y_{m}(\vK)\big]^2\big),   \\
    \sigma_{H} &= -e^3 B \sum_{m} \int_{BZ} \frac{d^2 K}{(2\pi)^2} \big(\partial_{\mu} f[\epsilon_m(\vK)]\big)
    \big( v^X_{m}(\vK) \tau_m (\vK) \big) \big[v^Y_{m}(\vK) \partial_{K_{X}} - v^X_{m}(\vK) \partial_{K_Y} \big] 
    \big( v^Y_{m}(\vK) \tau_m (\vK) \big), 
  \end{align}
\end{widetext}
where $m$ runs over the self-consistent bands, $\epsilon_m(\vK)$ is the energy of the $m$th band at momentum $\vK$,
$f[\epsilon_m(\vK)]$ is the corresponding Fermi-Dirac distribution, 
$v^i_m (\vK) = \partial_{K_i}\epsilon_m(\vK)$ is the corresponding $i$th component of the group velocity, 
and $\tau_m(\vK)$ is the corresponding momentum dependent scattering time. In this work we assume that the momentum dependence of the band
lifetimes arises from the momentum-dependent orbital character of the bands. Thus, we define $\tau_m$ to be
the weighted average of the orbital lifetimes,
\begin{align}
  \tau_m (\vec{K}) = \sum_{\sigma} &\tau_{\text{a}_{\text{1g}}} |\psi_m(\text{a}_{\text{1g}},\sigma,\vec{K})|^2 + \\
  \nonumber &\tau_{\text{e}_{\text{g}}^{\prime}} \big[|\psi_m(\text{e}_{\text{g}+}^{\prime},
  \sigma,\vec{K})|^2 + |\psi_m(\text{e}_{\text{g}-}^{\prime},\sigma,\vec{K})|^2 \big].
\end{align}

\begin{figure*}
\includegraphics[width=0.85\textwidth]{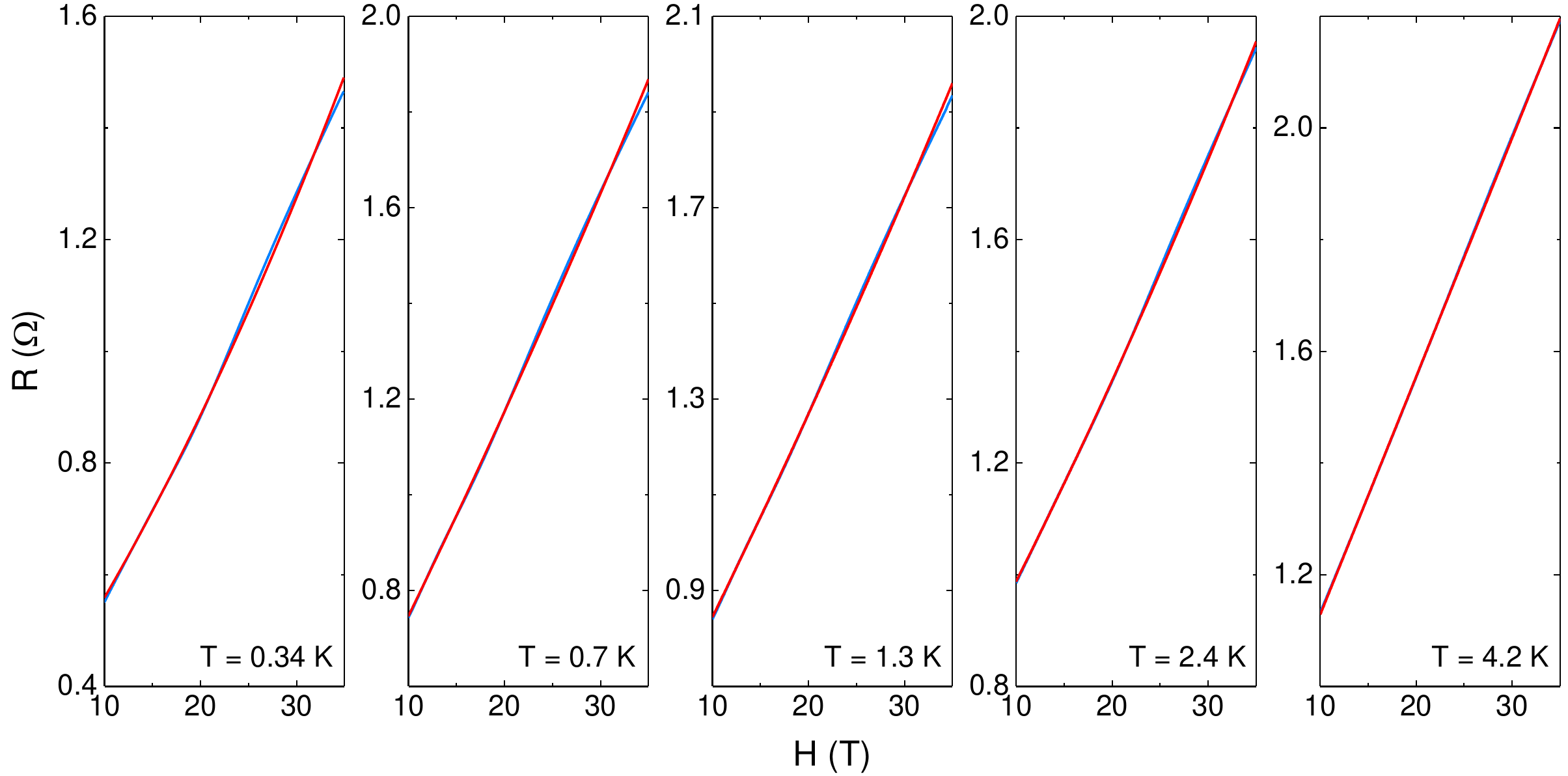} 
  \caption{Perpendicular magnetoresistance (blue) at $V_{\text{g}} =$ 6 V for $T=$ 0.34, 0.7, 1.3, 
  2.4, and 4.2 K, together with fits to a quadratic ($a_0 + a_{\text{1}}H + a_{\text{2}}H^2$) 
  function (red).}
\label{Fig-Supp-RH}
\end{figure*}

This allows $\tau_m(k)$ to trace the change in orbital character along the Fermi contour. As discussed in the
main text, experimental observations imply that the second band has a larger mobility than the first. 
Therefore we choose $\tau_{\text{e}_{\text{g}}^{\prime}} > \tau_{\text{a}_{\text{1g}}}$. We note that the orbital lifetimes
are assumed to be independent of momentum and energy, which is unlikely to be true in the real material. 
However, as shown in Fig. 3(e), our results are in agreement with the basic features observed experimentally.
Therefore, we believe that the essential physics is correctly captured by our model.

\subsubsection*{Incorporating Additional Layers}

The interface model can be easily extended to include additional layers, since the connectivity of the new atoms 
is given by the $k_3$ dependent terms in $H_{\text{0}} + H_{\text{NNN}}$ [Eqs.~(S27)$-$(S32)]. Fig.~\ref{Fig-Supp-6L} shows the 
results of the self-consistent calculation with six layers with the parameters that were used for Fig. 3 of 
the main text, except for a smaller confining potential, which allows for larger mixing with the deeper layers. 
Nevertheless, the behavior remains essentially the same as in the three-layer case.

\begin{figure}[b]
\includegraphics[width=0.9\columnwidth]{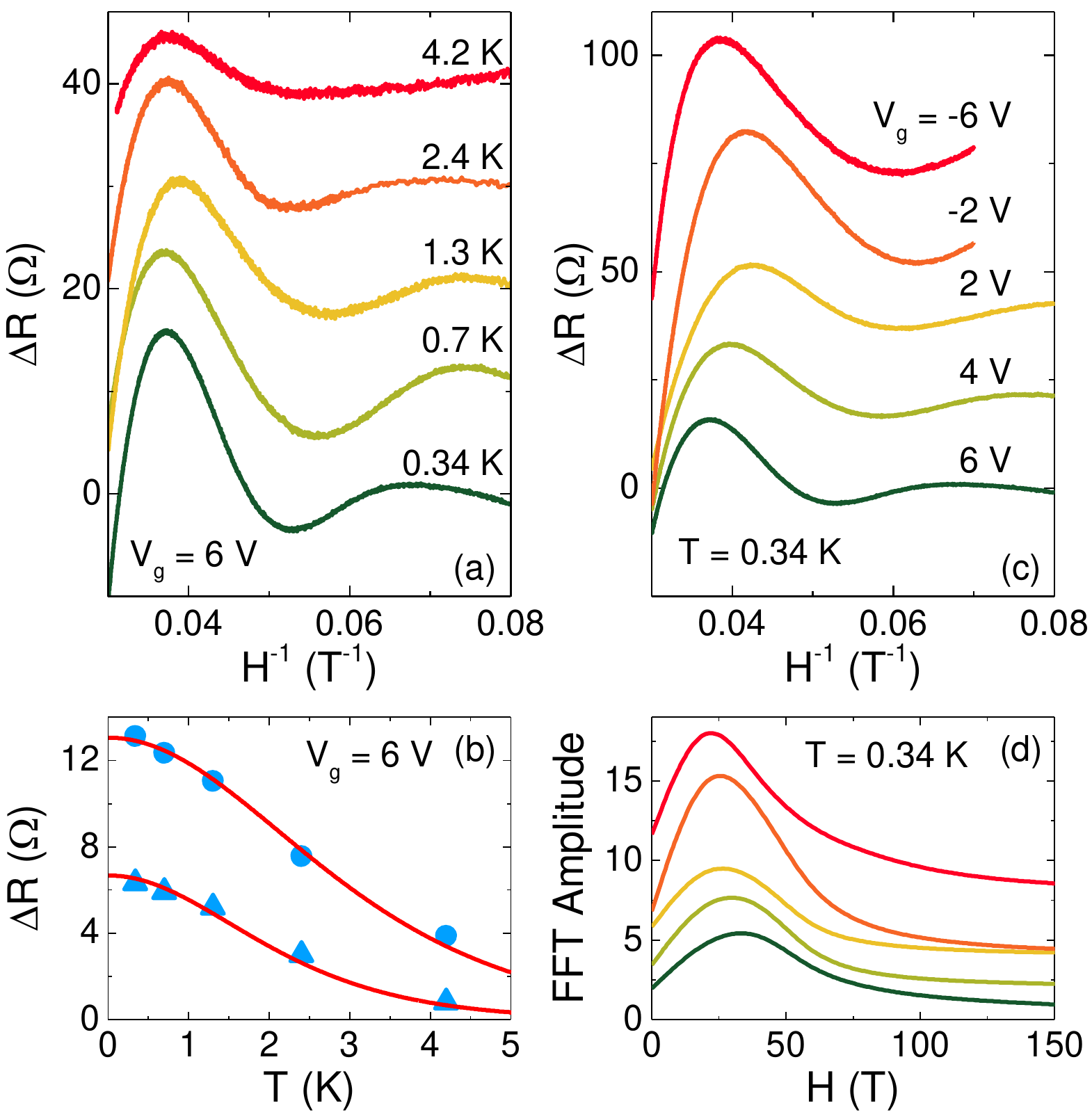} 
  \caption{(a) $\Delta R$ as a function of $H^{-1}$ for $T=$ 0.34, 0.7, 1.3, 2.4, 
    and 4.2 K at $V_{\text{g}}=$ 6 V. (b) The temperature dependence of $\Delta R$ at the
    first maxima and minima in (a). The red lines are the fits  
    to Eq.~(\ref{Eq-SdH}). (c) $\Delta R$ as a function of $H^{-1}$ for $V_{\text{g}}=$ 6, 4, 2, $-$2, 
    and $-$6 V at $T=$ 0.34 K. (d) The FFT amplitude of $\Delta R$ presented in (c).}
\label{Fig-Supp-SdH}
\end{figure}

\section{III.\,\,\,\,\,\,      Analysis of the Experimental \,\,\,\,\,\,\,\,\, Magnetotransport Data}

We measure the modification of the device resistance ($R$) due to a perpendicular magnetic field 
($H$) at $T=$ 340 mK for various $V_{\text{g}}$ (Fig.~\ref{Fig-Supp-RV}) and at $V_{\text{g}} =$ 
6 V for various temperatures (Fig.~\ref{Fig-Supp-RH}). All the magnetoresistance (MR) measurements 
show a strong positive MR as reported previously \cite{YoramAMR111}. In order to 
extract the Shubnikov-de Haas (SdH) signal, we fit $R$($H$) to a second order polynomial in $H$, 
$a_0 + a_{1} H + a_{2} H^{2}$ (See Figs. S3 and S4) and subtract the polynomial background from 
$R(H)$ to obtain $\Delta R$. The extracted oscillatory resistance $\Delta R$ is plotted in 
Fig.~\ref{Fig-Supp-SdH}(a) for different temperatures at $V_{\text{g}}=$ 6 V, and in Fig.~\ref{Fig-Supp-SdH}(c) 
for different $V_{\text{g}}$ at $T=$ 0.34 K. To further analyse $\Delta R$, 
we use the standard SdH expression \cite{shoenberg},
\begin{equation}
  {R_{\text{SdH}}}={R_{0}}{e^{-\alpha{T_{\text{D}}}/H}}\frac{{\alpha T/H}}{{\sinh(\alpha T/H)}}\sin(2\pi F/H),\label{Eq-SdH}
\end{equation}
where $R_{0}$ is a constant pre-factor, $\alpha=2\pi{m^{*}}{k_{B}}/\hbar e$, $m^{*}$ is the 
cyclotron effective band mass, 
$T_{\text{D}}$ is the Dingle temperature, and $F$ is the frequency of the oscillation. 
The best fits to above expression for the oscillation amplitude for the first maxima and 
minima yield $m^{*}=$ 1.6 $\pm$ 0.1 $m_{\text{e}}$ ($m_{\text{e}}$ being the electron mass) and 
$T_{\text{D}}=$ 5.4 K corresponding to $V_{\text{g}}=$ 6 V [Fig.~\ref{Fig-Supp-SdH}(b)]. 
By taking the fast Fourier transform (FFT) of $\Delta R$ for different $V_{\text{g}}$ [Fig.~\ref{Fig-Supp-SdH}(d)], 
we determine the SdH frequency $F$ as the peak field, which is related to the 
cross-sectional area $A({\epsilon_{\text{F}}})$ of the 2D Fermi line through the Onsager relation,
\begin{align}
  F=\frac{\hbar}{2\pi e} A({\epsilon_{\text{F}}}).
\end{align}

\begin{figure}[t]
\includegraphics[width=1.0\columnwidth]{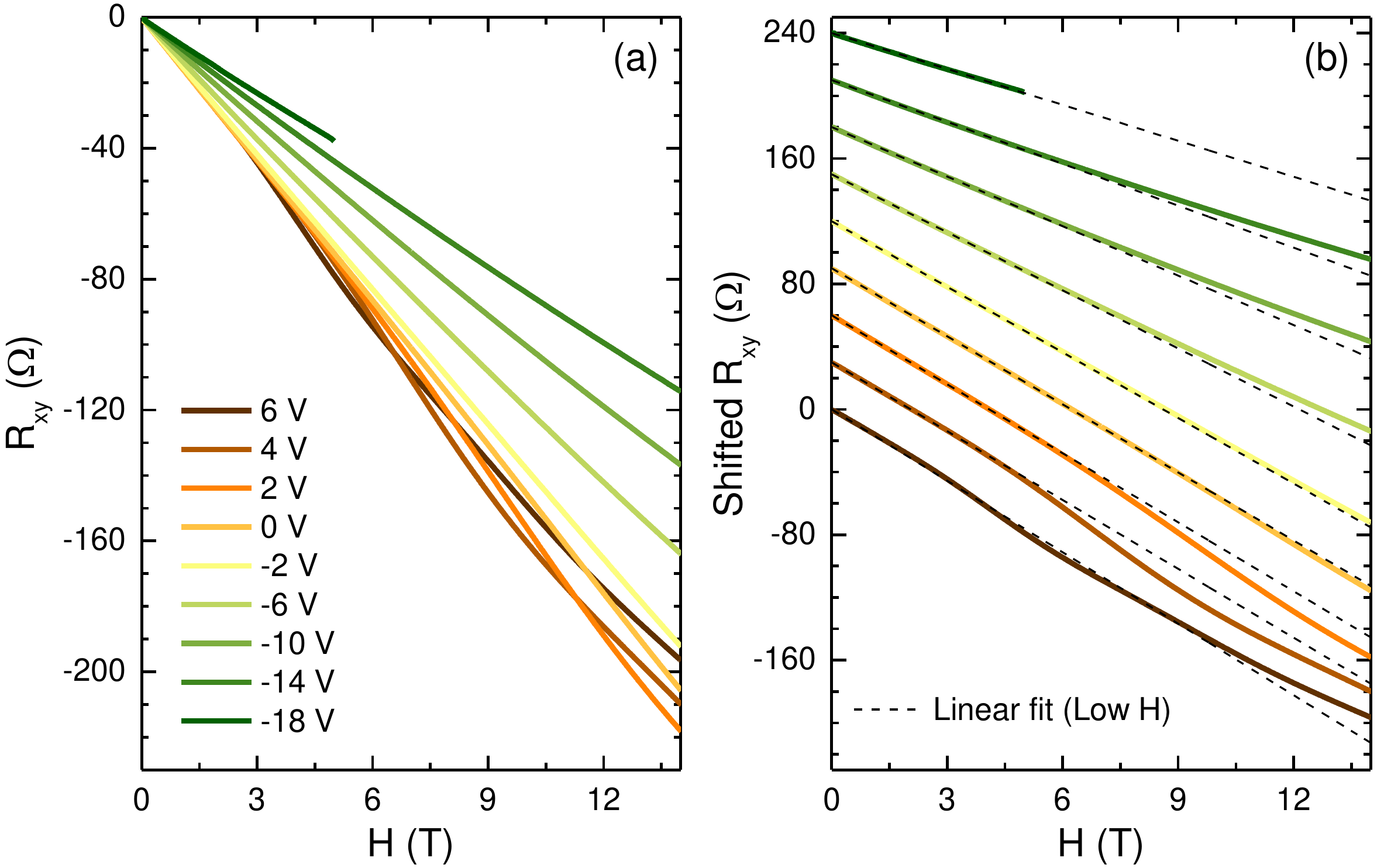} 
  \caption{(a) The Hall resistance ($R_{\text{xy}}$) as a function of perpendicular magnetic field ($H$) for various 
  $V_{\text{g}}$ at $T =$ 0.34 K. (b) $R_{\text{xy}}$ shifted by 30 $\Omega$ from each other for clarity. 
  The black dashed lines are the linear fit to low field Hall data. }
\label{Fig-Supp-Rxy}
\end{figure}

\noindent
Irrespective of the shape of the Fermi contour, this gives the sheet carrier density of the band 
which contributes to the SdH oscillations as $n_{\text{SdH}}={N_{\text{v}}}{N_{\text{s}}}eF/h$, 
where $N_{\text{v}}$ and $N_{\text{s}}$ 
are the number of valleys and spin species respectively. Fig.~1(a) of the main text presents $n_{\text{SdH}}$ calculated
for a single valley and $N_{\text{s}} = 2$.   
Fig.~\ref{Fig-Supp-Rxy} presents the low field Hall measurement performed on the sample at 0.34 K. 
For all $V_{\text{g}}$ we observe negative Hall slope, which implies the presence of electron-like 
charge carriers according to standard Drude model. We employed the Drude expressions to determine the 
Hall carrier density $n_{\text{Hall}}$ [$= (e|R_{\text{H}}|)^{-1}$, where $R_{\text{H}}$ is the Hall 
coefficient], which is presented in Fig. 1(a) of the main text.

\begin{figure}
\includegraphics[width=0.9\columnwidth]{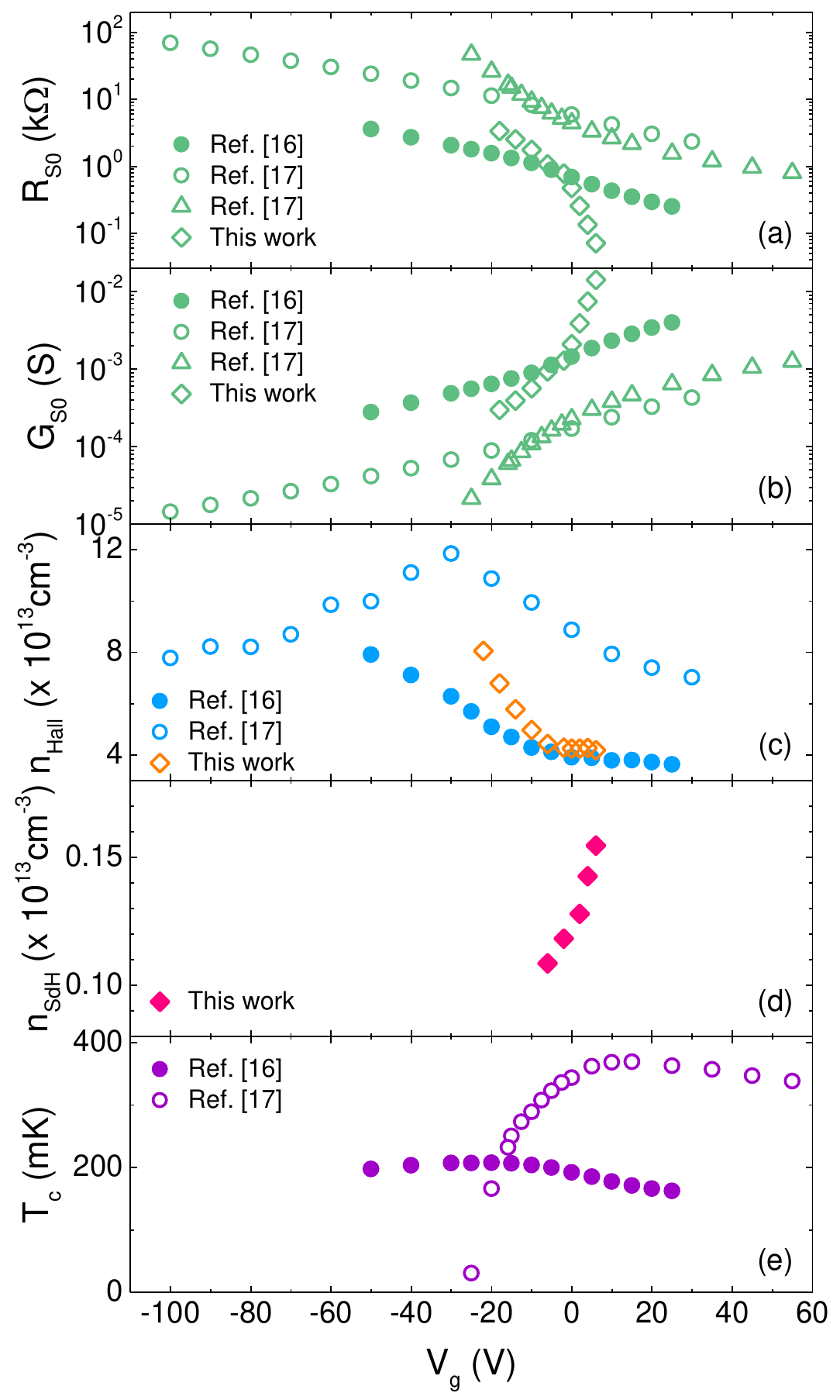} 
  \caption{ Gate dependence of (a) $R_{\text{S0}}$,(b) $G_{\text{S0}}$,(c) $n_{\text{Hall}}$, (d) $n_{\text{SdH}}$, 
  and (e) $T_{\text{c}}$. We present the new data as well as our previous measurements  on 
  $(111)$ \LAO/\STO~for comparison \cite{yoramSOSC,yoram1805}. Note that two independent measurements were 
  carried out in \cite{yoram1805} to determine $n_{\text{Hall}}$ and $T_{\text{c}}$, which leads to two sets 
  of $R_{\text{S0}}$ and $G_{\text{S0}}$ (as a function of $V_{\text{g}}$). }
\label{Fig-Supp-Raw}
\end{figure}

Figure \ref{Fig-Supp-Raw} shows the dependence on gate voltage ($V_{\text{g}}$) of the sheet resistance ($R_{\text{S}}$), 
sheet conductance ($G_{\text{S0}} = 1/R_{\text{S}}$), sheet carrier density determined from low-field Hall measurements 
($n_{\text{Hall}}$) and from quantum oscillations ($n_{\text{SdH}}$) as well as the superconducting critical 
temperature ($T_{\text{c}}$). Fig.~1 of the main text presents $n_{\text{Hall}}$, $n_{\text{SdH}}$, and $T_{\text{c}}$
as a function of $G_{\text{S0}}$.


\begin{thebibliography}{55}
\bibitem [{fnL()}]{fnLayers}%
  \BibitemOpen
  \href@noop {} {}\bibinfo {note} {Each additional layer adds new subbands and
  slightly modifies the structure of the lowest subband. At the densities
  relevant here, only the lowest subband plays an important role and we have
  verified that including additional layers (beyond 3) has an insignificant
  effect on our results~\cite{supp}.}\BibitemShut {Stop}%
\end{thebibliography}
\end{document}